\documentclass[prd,aps,12pt,amssymb,preprintnumbers,amsmath,amsfonts]{revtex4}
\usepackage{latexsym}
\usepackage[latin1]{inputenc}
\usepackage{pifont}
\usepackage{hyperref}
\newcommand{\ba}{\begin{eqnarray}}
\newcommand{\ea}{\end{eqnarray}}

\newcommand{\be}{\begin{equation}}
\newcommand{\ee}{\end{equation}}
\newcommand{\pa}{\partial}

\newcommand{\uu}{\slashed{u}}
\newcommand{\vv}{\slashed{v}}
\newcommand{\nn}{\nonumber}
\newcommand{\vt}{\slashed{\tilde{v}}}
\newcommand{\Dt}{\slashed{D}_\perp}

\newcommand{\K}[1]{{\bf K}_{#1}}

\newcommand{\ipp}[1]{{\int \frac{d^3{#1}}{(2\pi)^3}}}
\newcommand{\ippe}[1]{{\int_{#1} }}

\newcommand{\qq}{{\bf{q}}}

\newcommand{\qpar}{{q_\parallel}}
\newcommand{\qperp}{{q_\perp}}
\renewcommand{\Re}{\textrm{Re\hspace{0.5mm}}}

\usepackage{slashed}
\usepackage{dcolumn}
\usepackage{graphicx}
\usepackage{color}

\usepackage{ulem}
\definecolor{stcol}{rgb}{1,0,1}

\begin{document}
                                                                                
\date{\today}

\title{Chiral kinetic theory from the on-shell effective field theory: derivation of collision terms}

\author{Stefano Carignano}
\email{stefano.carignano@fqa.ub.edu}
\affiliation{Departament de F\'\i sica Qu\`antica i Astrof\'\i sica 
                   and Institut de Ci\`encies del Cosmos,
        Universitat de Barcelona,
        Mart\'\i $\,$ i Franqu\`es 1, 08028 Barcelona, Catalonia, Spain}

\author{Cristina Manuel}
\email{cmanuel@ice.csic.es}
\affiliation{Instituto de Ciencias del Espacio (ICE, CSIC) \\
C. Can Magrans s.n., 08193 Cerdanyola del Vall\`es, Catalonia, Spain
and \\
 Institut d'Estudis Espacials de Catalunya (IEEC) \\
 C. Gran Capit\`a 2-4, Ed. Nexus, 08034 Barcelona, Spain}
 
\author{Juan M. Torres-Rincon}
\email{torres-rincon@itp.uni-frankfurt.de}
\affiliation{Institut f\"ur Theoretische Physik, Johann Wolfgang Goethe-Universit\"at, Max-von-Laue-Strasse 1, D-60438 Frankfurt am Main, Germany}

\begin{abstract}
We show that the on-shell effective theory (OSEFT)  is the quantum field theory counterpart of a Foldy-Wouthuysen diagonalization of relativistic quantum mechanics for massless 
fermions. Thus,   it is  free of the {\it Zitterbewegung} oscillations that would yield an ill-defined meaning to the semiclassical transport approach at short distances if derived from the pure Dirac picture. We present a detailed derivation of the collision terms in the chiral kinetic theory using the OSEFT. Collision integrals are derived up to order $1/E$, where $E$ is the energy of an on-shell fermion. At this order, the collision terms depends on the spin tensor of the fermion, and in the presence of chiral imbalance, it describes how a massless fermion of a given helicity interacts differently with the transverse photons of different circular polarization. In order to back up our results,  we check that they allow us to reproduce the fermion decay rate in an ultradegenerate plasma with a chiral imbalance computed directly from QED.
\end{abstract}
\maketitle

\section{Introduction}

In this manuscript we continue our work on the derivation of  chiral transport theory   from the on-shell effective field theory (OSEFT)~\cite{Manuel:2014dza,Carignano:2018gqt}. We first discuss how the OSEFT is equivalent to a  Foldy-Wouthuysen  diagonalization for massless fermions, and discuss several of its subtleties. Then we present a detailed derivation of the collision terms of the chiral kinetic theory, as derived from OSEFT. We assume a system composed of massless chiral charged fermions interacting through electromagnetic fields.

The chiral kinetic theory (CKT) was first proposed in Refs.~\cite{Son:2012wh,Son:2012zy,Stephanov:2012ki} starting with the action of a point particle modified by the Berry curvature, together with a modified Poisson bracket structure. Different alternative derivations have been discussed in the literature since then~\cite{Chen:2012ca,Manuel:2013zaa,Manuel:2014dza,Hidaka:2016yjf,Gorbar:2016ygi,Hidaka:2017auj,Mueller:2017lzw,Mueller:2017arw,Hidaka:2018ekt,Gorbar:2017awz,Huang:2018wdl,Gao:2018wmr}. 

The CKT was initially formulated as a sort of semi-classical approach, where the point particle picture of classical mechanics could be used to see how some relevant quantum effects modify different transport phenomena. It is also possible to derive such framework from quantum field theory.  However, one could run into the same sort of ambiguities as those that were found when trying to give a semi-classical interpretation to the Dirac equation. It is well known that the Dirac Hamiltonian mixes up the dynamical evolution of positive and negative energy solutions~\cite{Foldy:1949wa,Thaller}, an effect that is apparent in the famous {\it Zitterbewegung} (ZB) motion of relativistic electrons~\cite{Schroedinger},  which occurs over distances of the order of the  Compton wavelength of the particle, $\lambda^c = \hbar/(mc)$.  The ZB oscillations also have an effect on the relativistic transport approach~\cite{BialynickiBirula:1991tx}. To eliminate the ZB oscillations one should  fold the quantum Wigner function with a coarse graining function, such that the ZB oscillations are averaged out~\cite{Rafelski:1993uh}. These steps are also needed to render a proper classical probabilistic meaning to the quantum approach, as otherwise the quantum Wigner function $S$ can take negative values at short distances. Typically these considerations are overlooked, as one considers that the quantum transport approach is only valid at enough large scales, much larger than $\lambda^c$, by imposing as a condition $| \lambda^c  \gamma_\mu \pa^\mu_X S(X,p)| \ll |S(X,p)|$. In the massless case, this condition  is also used by simply replacing $\lambda^c$ by $\lambda^{\rm dB}= \hbar c/E$,  the de Broglie wavelength, where $E= cp$ is the fermion energy~\cite{relativisticbook}.

It is possible to get rid of the ZB oscillations of the quantum relativistic fermions by performing a Foldy-Wouthuysen (FW) diagonalization of the Dirac Hamiltonian~\cite{Foldy:1949wa}. The physical interpretation of the rotated fields differs from those of the original Pauli-Dirac picture (see Sec.~\ref{sec:discussion} for a more detailed discussion). The pioneering work of FW was done for massive relativistic fermions, and showed that one could properly disentangle the particle and antiparticle sectors interacting with weak electromagnetic fields as an expansion in $1/m$, where $m$ is the fermion mass. Later on, it was shown that the same could be achieved with the use of effective field theories, such as non-relativistic quantum electrodynamics (NRQED)~\cite{Caswell:1985ui}.

In a similar fashion, it is also possible to disentangle the dynamics of  particle and antiparticle sectors of massless relativistic fermions, as an expansion in $1/E$,  assuming that the energy is the large scale of the problem.  We have derived an effective field theory approach for that purpose~\cite{Manuel:2014dza}. A detailed derivation of the OSEFT Lagrangian has been given in Refs.~\cite{Manuel:2014dza,Manuel:2016wqs,Manuel:2016cit,Carignano:2018gqt}. Let us also mention here that the OSEFT has been used also for other purposes not related to transport theory, such as describing the power corrections of the hard thermal loop amplitudes of QED~\cite{Manuel:2016wqs,Manuel:2016cit}, providing the same result as that derived directly from QED~\cite{Carignano:2017ovz}.

In this manuscript we   prove that the OSEFT Lagrangian can be derived order by order via successive FW transformations of the QED Lagrangian. Once noted this equivalence, several implications appear in the interpretation of our fermion fields, as the OSEFT  ``particle field'' is a combination of the original particle and antiparticle components of the Dirac spinor, as it occurs after a FW diagonalization. In other words, the effective semi-classical  particle is seen as a combination of Dirac particles and antiparticles, and has different properties~\cite{Costella:1995gt}.

All the above discussion is relevant to understand the differences of the chiral transport equation as derived from the OSEFT, or as derived from the Dirac picture, after an $\hbar$-expansion~\cite{Hidaka:2016yjf}. It was noted in Ref.~\cite{Lin:2019ytz} that these differences arise because the two equations act on different degrees of freedom, which is correct. However, while the OSEFT is free of the ZB effects, a  systematic procedure to eliminate the ZB oscillations as derived from the Dirac picture should be worked out, such as the coarse graining mentioned in Ref.~\cite{Rafelski:1993uh}, that would necessarily introduce a minimal length scale in the resulting framework. Details of this procedure, and how it might affect the resulting transport approach when going beyond the pure classical limit, are yet to be discussed. This is a relevant question in CKT as the transport equation contains terms of the order $\sim \pa_\mu^X/E$ which could measure fluctuations of the Wigner function at scales that are sensitive to the ZB effects.

This paper is structured as follows. In Sec.~\ref{FW-section} we show that the OSEFT Lagrangian up to order $1/E^2$, which was used in Ref.~\cite{Carignano:2018gqt} for the derivation of the  collisionless CKT, can be recovered from the QED Lagrangian  by carrying out subsequent FW diagonalizations. Higher-order terms could be derived as well, but we leave this for future projects.  In Sec.~\ref{sec:discussion} we provide several discussions related to the FW versus Pauli-Dirac representations, reparametrization invariance of the OSEFT,  and the so-called side-jumps. In Sec.~\ref{Sec-GeneralTransport} we present the basic steps to construct the transport equation from quantum field theory. This is a reminder of the standard techniques, based mostly on Refs.~\cite{Blaizot:1999xk,Blaizot:2001nr} for QED,   and could be skipped by those readers who are familiar with them. In Sec.~\ref{sec:transpOSEFT} we comment on the differences and particularities in the derivation of the transport equation in the OSEFT case. In Sec.~\ref{sec:colloseft} we provide the basic steps on the calculation of the collision terms in the OSEFT. In Sec.~\ref{sec:decayrate} we compute the fermion decay rate in an ultradegenerate plasma using kinetic theory with a QED matrix element, and check that after a $1/E$ expansion it can be reproduced using our OSEFT results for an isotropic plasma. Finally we present our conclusions in Sec.~\ref{sec:conclusions}. We devote App.~\ref{app:notation}  to collect some of the notation we use in this article.  In App.~\ref{app-photon} we recall the structure of the photon propagator in a medium with chiral imbalance.  We will use natural units in this paper $\hbar = c= k_B= 1$, unless otherwise indicated.

\section{Derivation of the OSEFT from a FW diagonalization}
\label{FW-section}

Consider a massless charged fermion in a given frame with energy $E$ and
light-like velocity $v^\mu = (1 , {\bf v})$, where ${\bf v}$ is a unit 3-vector. Let us 
define ${\tilde v}^\mu= (1 , -{\bf v}) $, which is also a light-like vector. Thus $v^2 = { \tilde v}^2=0$, but $v \cdot {\tilde v} =2$. We also define the orthogonal projector,
\be 
\label{T-projector}
P^{\mu \nu}_\perp = g^{\mu \nu} - \frac 12 \left( v^\mu {\tilde v}^\nu +{\tilde v}^\mu v^\nu \right) \ .
\ee 
 The massless QED Lagrangian describing this fermion reads
\be {\cal L}= \bar{\psi}^{(0)} i \slashed{D} \psi^{(0)} \ , \ee
where $\psi^{(0)}$ is the standard Dirac spinor, and $D^\mu=\pa^\mu + ieA^\mu$ is the covariant derivative.

We first perform the change in the field
\be \psi^{(1)} = \exp \left( i E v\cdot x\right)   \psi^{(0)} \ , \ee
and using the decomposition (\ref{T-projector}) the Lagrangian can be written as
\be 
{\cal L} = {\bar \psi}^{(1)}   \left( i \Dt + \frac{\vt}{2}  (i  v\cdot D) + \frac{\vv}{2}  \left( 2E +i \tilde{v} \cdot D \right) \right) \psi^{(1)} \ ,
\label{Initial-L}
\ee
where $\Dt =\gamma_\perp \cdot D=P_\perp^{\mu \nu} \gamma_\mu D_\nu$.

We define particle and antiparticle projectors as
\be \label{eq:projectors}
P_v = \frac12 \vv \uu \ , \qquad P_{\tilde{v}} = \frac12 \vt \uu \ ,
\ee
respectively, where $u^\mu = (1, {\bf 0})$ is a 4-vector describing the rest frame. By noting that $\vv P_v = \vt P_{\tilde{v}} = 0 $, $\vt P_v =2 \uu P_v$,  and $  \vv P_{\tilde{v}}= 2 \uu  P_{\tilde{v}}$,  one can check that Eq.~(\ref{Initial-L}) reproduces the Lagrangian Eq.~(55) of Ref.~\cite{Manuel:2014dza} for a single fermion of energy $E$ and velocity $v^\mu$.

Unfortunately, Eq.~(\ref{Initial-L}) mixes up particle and antiparticle degrees of freedom due to the presence of the ``odd'' operator $i\Dt$.  
To disentangle these two degrees of freedom of the Dirac field a couple of different techniques were used in Ref.~\cite{Manuel:2014dza}. First, a FW diagonalization
at the Hamiltonian level, performed as an expansion in $\hbar$. An effective field theory, the OSEFT, was then also proposed to separate particle and antiparticle degrees
of freedom of the Dirac field at a quantum field theory level. While it is not {\it a priori} obvious, the two approaches are fully equivalent. To show this we present here a third equivalent way, which consists of performing a FW diagonalization at the Lagrangian level, which  allows us to recover the OSEFT Lagrangian at a given order of accuracy. These three techniques
have been proven to be equivalent for relativistic massive fermions, when the diagonalization is carried out as an expansion in $1/m$ \cite{Korner:1991kf,Holstein}, the inverse of the
fermion mass.

In order to be fully general, and to recover the results of OSEFT in an arbitrary frame~\cite{Carignano:2018gqt}, from now on we will allow the frame vector $u^\mu$ to be an arbitrary time-like vector $u^2 =1$ fulfilling  the condition
\be \label{eq:refframe}
u^\mu = \frac{v^\mu + {\tilde v}^\mu}{2} \ .
\ee

To remove the odd operator  in Eq.~(\ref{Initial-L}) we carry out the canonical transformation 
\be \psi^{(2)} = \exp\left({ \frac{\uu \,{\cal S}^{(1)}}{2E} }\right)  \psi^{(1)} \ ,  \qquad  {\cal S}^{(1)} \equiv i  \Dt \ .
\ee
The Lagrangian acting on the new field reads
\begin{align} 
{\cal L} &= {\bar \psi}^{(2)} \exp \left(  \frac{\uu i\Dt}{2E} \right)  \left(  i \Dt + \frac{\vt}{2}  (i  v\cdot D) + \frac{\vv}{2}  \left( 2E +i \tilde{v} \cdot D \right)   \right) \exp \left( - \frac{\uu i\Dt}{2E} \right) \psi^{(2)} \ .
\end{align}
Using the formula 

\be e^{A} B e^{-A} =B + [A,B] + \frac{1}{2!} [A,[A,B]] + \frac{1}{3!} [A,[A,[A,B]]] + \cdots 
\ee
one can explicitly work out every term in the Lagrangian in terms of a $1/E$ expansion
\begin{align}
\label{Second-L}
{\cal L} &= \bar{\psi}^{(2)} \left(    \frac{\vt}{2}  (i  v\cdot D) + \frac{\vv}{2}  \left( 2E +i \tilde{v} \cdot D \right)  \right) \psi^{(2)} \\
& +  \frac{1}{2E} \bar{\psi}^{(2)} \left( (i \Dt)^2 \uu -   i \Dt  i v\cdot D \frac{\vv \vt}{4} -   i  v\cdot D i \Dt \frac{\vt \vv}{4}
-   i \Dt i \tilde{v}\cdot D \frac{\vt \vv}{4} -  i   \tilde{v}\cdot D i \Dt \frac{\vv \vt}{4} 
\right) \psi^{(2)}
 \nn \\
& -\frac{1}{16E^2} \bar{\psi}^{(2)}  \Big(  \left \{(i\Dt)^2, i v\cdot D \right \} \vt +2 i\Dt i v\cdot D i\Dt \vv + 2 i\Dt i \tilde{v}\cdot D i\Dt \vt
 \nn \\
 & 
+ \left\{ (i\Dt)^2, i\tilde{v}\cdot D \right \} \vv     \Big)  \psi^{(2)}
 - \bar{\psi}^{(2)} \frac{1}{3E^2} (i\Dt)^3  \psi^{(2)}+ {\cal O} \left(\frac{1}{E^3} \right) \ .
 \nn
\end{align}

Notice that in Eq.~(\ref{Second-L}) we have eliminated the ``odd'' operator at leading order, but additional odd operators connecting particles and antiparticles at orders $1/E$ and $1/E^2$ are still present. To eliminate those at ${\cal O} (1/E)$ we need to carry out an additional transformation,
\be \psi^{(3)} = \exp \left( \frac{\uu \, {\cal S}^{(2)}}{2E} \right) \psi^{(2)}  \ ,
\ee
with
\be {\cal S}^{(2)} \equiv - \frac{1}{2 E} \left[  (i \Dt  i v\cdot D + i   \tilde{v}\cdot D i \Dt)P_v + (i  v\cdot D i \Dt 
+ i \Dt  i \tilde{v}\cdot D ) P_{\tilde v}
\right] \ .
 \ee
This transformation generates itself new odd terms at subleading orders, while keeping the even operators untouched. Yet another transformation, 
\be \psi^{(4)} = \exp \left( \frac{\uu \, {\cal S}^{(3)}}{2E} \right) \psi^{(3)}
\ , 
\ee
with
\begin{align}
{\cal S}^{(3)} & \equiv   \frac{1}{4E^2}\left( i\Dt (i v\cdot D)^2 + 2 i
\tilde{v}\cdot D i\Dt i v\cdot D+ (i \tilde{v}\cdot D)^2 i\Dt  \right)
P_v  \ \\
& +  \frac{1}{4E^2}\left( i\Dt (i \tilde{v}\cdot D)^2 + 2 i v\cdot D
i\Dt i \tilde{v}\cdot D+ (i v\cdot D)2 i\Dt  \right) P_{\tilde{v}}
  \nn 
-  \frac{1}{3E^2} (i\Dt)^3 
\end{align}
will remove all the pieces that mix particles and antiparticles at order $1/E^2$ (while there will be odd operators at the following orders in the energy expansion). Successive canonical transformations should be done at every order in the energy expansion to achieve a full diagonalization.

It is now easy to see how these FW partial diagonalizations allow us to reproduce the OSEFT Lagrangian at a certain order of accuracy. If we define the particle/antiparticle components at a given order $(n)$ in the FW diagonalizations:
 \be
\chi^{(n)} \equiv  P_v \psi^{(n)}  \ , \qquad \xi^{(n)} \equiv P_{\tilde v} \psi^{(n)}
\ee
then at ${\cal O}(1/E^2)$ the particle Lagrangian reads 

\be
{\cal L} = \bar{\chi}^{(4)} \left(   i v\cdot D  + \frac{1}{2E} (i\Dt)^2  -\frac{1}{8E^2} \left \{(i\Dt)^2, i v\cdot D \right \} 
 - \frac{1}{4E^2} i\Dt i\tilde{v}\cdot D i\Dt  \right) \frac {\vt}{2} \chi^{(4)}  \ ,
\label{eq:L4}
\ee
plus the analogous term for the antiparticle field $\xi^{(4)}$.

We stress that the diagonalization we have carried out for massless fermions assumes that $E$ is the hard scale, larger than the values of the electromagnetic fields and their gradients, and also of the derivatives of the Dirac field.
Note also that
\be
\chi^{(4)} = e^{i E v \cdot x} \left ( \chi^{(0)} + \frac{ \uu i \Dt}{E} \xi^{(0)} - \frac{ (i \Dt)^2}{8 E^2} \chi^{(0)} - \frac{\uu}{4 E^2} (i  v\cdot D i \Dt 
+ i \Dt  i \tilde{v}\cdot D ) \xi^{(0)} 
\right)  + {\cal O} \left(\frac{1}{E^3} \right) \ ,
\ee
that is, the new particle field is a combination of the particle and antiparticle fields of the original Dirac picture [which is order $(0)$]~\cite{Foldy:1949wa}. The covariant derivatives in the expansion also tell us about the non-local relation between the original Dirac picture and the FW one.

Let us note that the Lagrangian (\ref{eq:L4}) contains temporal derivatives beyond the leading order term. Exactly as in Ref.~\cite{Manuel:2016wqs}, we perform a local field redefinition to eliminate these. Thus, after doing
\be
\label{LFR}
{\tilde \chi}  \equiv \left( 1 + \frac{(i\Dt)^2\uu}{4 E^2} \right)  \chi^{(4)} \ , \\
\ee
in the Lagrangian~(\ref{eq:L4}) we end up at order $1/E^2$

\be
{\cal L} = \bar{\tilde \chi} \left(   i v\cdot D  + \frac{1}{2E} (i\Dt)^2  
 +\frac{1}{8E^2} \left( \left\{ (i\Dt)^2, (iv\cdot D - i \tilde{v} \cdot D)  \right\}  -    [i\Dt, [i\tilde{v}\cdot D ,i\Dt]] 
 \right)\right) \frac{\vt}{2} {\tilde \chi} \ ,
\ee
which is the OSEFT Lagrangian deduced in Ref.~\cite{Manuel:2016wqs}, and used in Ref.~\cite{Carignano:2018gqt} for the derivation of the chiral transport equation. 

While here we have shown how to derive the OSEFT Lagrangian associated with a single fermion, it is possible to generalize the method and perform the diagonalizations
associated with having several fermions. One can also perform similar diagonalizations to derive the OSEFT Lagrangian for the on-shell antiparticles, simply 
exchanging $E \rightarrow -E$, and $v^\mu \leftrightarrow {\tilde v}^\mu$ ~\cite{Carignano:2018gqt} in all the preceding equations.

In Ref.~\cite{Manuel:2014dza} the OSEFT was derived using the modern language of effective field theories, where to describe on-shell particles one integrates out the off-shell modes.  Like in the QED Lagrangian these two set of modes are inherently coupled through the equations of motion. When the off-shell components are integrated out, only particles remain in the effective theory at the expense of having an infinite series of operators in the Lagrangian, but suppressed by successive powers of $1/E$. The FW diagonalization allows for a similar decoupling of particles and antiparticles order by order in $1/E$.  
As we arrive at the same result with FW diagonalizations,  we can therefore conclude that this and the  original  OSEFT approach describe the same physics.

\section{Discussion: OSEFT/FW picture and side jumps~\label{sec:discussion}}

As our formulation of the CKT is based on the OSEFT, which corresponds to the FW picture of relativistic quantum mechanics, we review in this section some of
its peculiarities. We also recall how the side jumps of CKT first discussed in~\cite{Chen:2014cla,Chen:2015gta} can be recovered in the OSEFT, as shown in Ref.~\cite{Carignano:2018gqt}.

The distinction between the Pauli representation and Foldy-Wouthuysen's was discussed by Foldy himself~\cite{Foldy:1958zz}, and by Newton and Wigner~\cite{Newton:1949cq}. A summary of the differences are given in Table~\ref{tab:reps}. Some of them are crucial to understand the different approach to fermion collisions. 

\begin{table*}[ht]
\begin{center}
\begin{tabular}{|c|c|c|}
\hline
 & FW representation (OSEFT) & Pauli-Dirac representation \\ 
\hline
\hline
Electromagnetic interaction & Non-minimal coupling & Minimal coupling \\
Magnetic moment & \ding{51} & \ding{55} \\
\hline
Fermion structure & Spatial extent $\sim\frac{1}{E}$ & Pointlike \\
Position operator & Mean-position of the wave-package & Fermion position \\
Velocity operator & Darwin interaction & {\it Zitterbewebung} \\
\hline
Spin, $S$ & Conserved & Not conserved (only $J$ is) \\
Classical limit for $S$ & \ding{51} & \ding{55} \\
\hline
\end{tabular}
\caption{Summary of some properties of the different fermion representations, Foldy-Wouthuysen (equivalently in the OSEFT) versus Pauli-Dirac.~\label{tab:reps}}
\end{center}
\end{table*}

The first difference concerns how the electromagnetic interaction is coupled to fermions. In the Dirac Lagrangian it couples in a minimal way, via the covariant derivative $D_x^\mu = \pa_x^\mu + i eA^\mu(x)$. The use of the covariant derivative implies the use of kinetic momentum instead of the canonical one. In the resulting FW formulation (either nonrelativistic~\cite{Foldy:1949wa} or for massless fermions) it is explicit that some terms involving electromagnetic fields cannot be written in terms of a simple minimal coupling, e.g. there is a magnetic moment, or Pauli term, which couples like  $\sigma^{\mu \nu} F_{\mu\nu}$ in the OSEFT Lagrangian.

Another difference concerns the interpretation of the position operator. In the Pauli-Dirac representation one works with punctual fermions with a position indicated by the eigenvalues of the operator $x$. However, the inherent mixture between particles and antiparticles degrees of freedom of the Dirac equation makes the would-be velocity operator unusual. It has $\pm c$, where $c$ is the speed of light, as unique eigenvalues, and its equation of motion gives rise to the famous {\it Zitterbewegung} in which the trajectory has an oscillating motion~\cite{Thaller}. In the FW, on the contrary, the fermion has a spatial extent of the order of  $\sim 1/E$ ( or $1/m$  in the large mass limit), and the position operator refers to the average position associated with the ZB motion in the Dirac picture. There is no ZB, because positive and negative-energy modes are decoupled, and the velocity operator (also denoted as Newton-Wigner velocity~\cite{Newton:1949cq, Costella:1995gt}) has  continuum eigenvalues.
Another residual feature of the ZB appears in this formulation, the Darwin term, which can be interpreted as the interaction of the electromagnetic field with the spherical charge distribution of the fermion due to the ZB oscillations in position~\cite{Foldy:1952,Itzykson}.

Finally the spin operator is also differently interpreted. In the Pauli-Dirac representation it is not conserved as only the total angular momentum ${\bf J}={\bf S}+{\bf L}$ is. In addition, a semiclassical interpretation of the spin in Dirac representation is not immediate. In the FW picture the spin is conserved, independently of the angular momentum. Therefore, it has a well-defined classical limit.

Our effective field theory description of the classical transport approach also allows us to understand the behavior of CKT under Lorentz transformations. The side jumps of CKT can be understood as a consequence of the so called reparametrization invariance of OSEFT.

The OSEFT fields are labelled by the velocity $v^\mu$ and $E$, given in one specific frame. The frame vector $u^\mu$  is a time-like vector, and it is defined such that $u \cdot v =1$.
The OSEFT fields have a dependence on the residual momentum, with values much less than $E$. In turn, this implies that the OSEFT fields describe large distances. However, there is a redundancy in the theory, as small shifts in the velocity could be reabsorbed in the definition of the residual momentum, leaving the physics unchanged. This is the so-called reparametrization invariance (RI). On the other hand, an explicit choice of the vectors $v^\mu$ and $u^\mu$ seem to imply an apparent breaking of the Lorentz symmetry of the initial theory, QED, which does not have a dependence on any explicit vector.

In Ref.~\cite{Carignano:2018gqt} we showed that the OSEFT Lagrangian is reparametrization invariant, the symmetry can also be studied in a $1/E$ expansion.
The proof goes in parallel  to that of a different effective field theory for massless fermions, the soft-collinear effective field theory~\cite{Manohar:2002fd}. There are three types of RI symmetries, namely type I, type II and type III, that change infinitesimally the value of the vectors $v^\mu$  and ${\tilde v^\mu}$, without changing their light-like behavior, and preserving the property $u \cdot v =1$. The three types of RI transformations can also be interpreted in terms of combinations of rotations and Lorentz boosts (see also~\cite{Heinonen:2012km}).

It is clear that changes in the vectors $v^\mu$ and ${\tilde v^\mu}$ have to be accompanied by  redefinitions of the residual momentum. We listed in  Ref.~\cite{Carignano:2018gqt} the changes on the residual momentum after the three different types of RI symmetries. As the OSEFT fields have a dependence on the residual momentum, this implies that covariant derivatives acting on these fields also change under RI. As the Wigner function of chiral kinetic theory is constructed from the OSEFT fields, then we could show that under one specific type of RI transformation the distribution function of chiral kinetic theory
also changes, giving rise to the side jump effect first discussed in Ref.~\cite{Chen:2014cla}.

Reparametrization invariance is intimately linked to Lorentz invariance. One could think of preforming a Lorentz transformation that would change the values of $v^\mu, u^\mu$. Then, the physics described  by the OSEFT theory (and the subsequent CKT) would not be the same if the residual momentum, and thus the $x$ dependence of the OSEFT functions are not changed as prescribed by RI.

Summarizing, by showing that the OSEFT is RI invariant one can show that it is Lorentz invariant. The separation of scales implicit in OSEFT, and ultimately in the CKT, is finally responsible that functions that describe long distance behavior might change in a non-standard way under Lorentz symmetry. 

Let us also mention that the FW interpretation of the chiral transport theory gives a plausible explanation of some of the apparent paradoxes found in Ref.~\cite{Chen:2015gta}. In particular, in that reference it was found that in binary collisions of relativistic massless fermions there could be discontinuous jumps in their trajectories, when observed in different frames, if conservation of angular momentum was considered.
However, in the semi-classical description, one has to take into account the finite size of the fermions, and the relevant operators which act as conserved quantities.
It has been noted that the center of mass of extended spinning objects is frame dependent, and suffers a side jump when observed in a different frame~\cite{Costa:2011zn}. For 
relativistic massless fermion the mean position, which is the center of a charge distribution in the Dirac picture, also suffers from side jumps. These jumps described in 
Ref.~\cite{Chen:2015gta} would be interpreted in our formalism, not as discontinuous trajectories, but as displacements of the mean position of the effective fermion in binary collisions when observed in different frames.

However, we note that when deriving the CKT from quantum field theory the collision terms are not described in terms of classical trajectories. We will use the quantum field theory formulation in this article instead.

\section{Transport equations derived from quantum field theory}
\label{Sec-GeneralTransport}

In this section we give a very brief account on how transport equations can be derived from a fermionic quantum field theory. We basically follow here the formulation of Refs.~\cite{Danielewicz:1982kk,Elze:1986qd,Vasak:1987um,Botermans:1990qi,Blaizot:1999xk,Blaizot:2001nr,Carrington:2004tm} keeping the Dirac structure of the equations. We first ignore the effects of the electromagnetic fields, and finally comment on the pertinent modifications to maintain gauge invariance when considering them.

We use the real time formalism of thermal field theory. If 
$\psi$ denotes the Dirac field, the associated two-point function in the closed time path contour is given 
 by a $2 \times 2$ matrix~\cite{Danielewicz:1982kk}
\be
\label{progaKS}
{\cal S} (x,y) = 
\left( \begin{array}{cc}
  S^{c} (x,y) & S^< (x,y) \\
 S^>(x,y) & S^{a} (x,y) \\
\end{array} 
\right ) =
\left( \begin{array}{cc}
\langle T  \psi (x) \bar \psi (y) \rangle & - \langle \bar \psi (y)  \psi (x) \rangle \\
\langle   \psi(x) \bar \psi(y) \rangle & \langle {\tilde T}  \psi (x) \bar \psi(y) \rangle 
\end{array} 
\right ) \ ,
\ee
where $T$ denotes time ordering, and ${\tilde T}$ anti-time ordering. From the above functions, one can additionally define the retarded and advanced propagators as
\begin{align}
S^R (x,y) & =  i\theta(x_0-y_0) \ ( S^{>} (x,y)- S^{<} (x,y)) \ , \label{eq:Sret} \\
S^A (x,y) & =  -i\theta(y_0-x_0) \ ( S^{>} (x,y)- S^{<} (x,y)) \ , \label{eq:Sadv}
\end{align}
where $\theta$ denotes the step function. Note that not all Green functions are independent as 
\be  S^R(x,y)-S^A(x,y)=S^{>}(x,y)-S^{<}(x,y) \ . \ee

One can deduce  equations of motion for every component of the two-point Green function from the Kadanoff-Baym equations~\cite{Kadanoff-Baym}, including effects of the fermion self-energy $\Sigma(x,y)$.  Similar to~\cite{Blaizot:1999xk} we denote by $\Sigma^{\rm t}(x)$  a possible tadpole contribution.
The two-point fermion self-energy is also expressed as a $2 \times 2$ matrix, whose components obey the same relations as the components of the fermion propagator.
More particularly, the equation for $S^<(x,y)$ is~\cite{Danielewicz:1982kk,Elze:1986qd}
\be [ S_{0,x}^{-1} - \Sigma^{\rm t} (x) ] S^< (x,y) = (\Sigma^R \otimes S^< )(x,y) + (\Sigma^<  \otimes S^A) (x,y) \, \label{eq:KB1} \ ,
 \ee
where $S_{0,x}^{-1}$ is the inverse of the free propagator, and 
we have defined the convolution operator
\be
\label{convolution}  
(A \otimes B) (x,y) \equiv \int_{-\infty}^\infty d^4z  A (x,z) B (z,y) \ . \ee
One should also consider the Hermitian conjugate equation acting on $y$:
\be S^< (x,y) [S_{0,y}^{-1} - \Sigma^{\rm t} (y) ]^\dag =  (S^< \otimes  \Sigma^A) (x,y) +  ( S^R \otimes \Sigma^<)  (x,y) \ , \label{eq:KB2} \ee
where the operator $S_{0,y}^{-1,\dag}$ acts to the left, and we use that $\Sigma^{\rm t}  (y)$ is real.

The transport equation is obtained by considering the difference of the two equations. After some trivial manipulations, one can finally obtain 
\begin{align}
\label{diffe-eq}
& [ S_{0}^{-1} - \Sigma^{\rm t} , S^<] (x,y) - [ \Re \Sigma^R  \stackrel{\otimes}{,} S^<] (x,y) -  [ \Sigma^< \stackrel{\otimes}{,} \Re S^R ] (x,y)\nn \\
& =  \frac{i}{2} \{ \Sigma^> \stackrel{\otimes}{,} S^< \} (x,y) - \frac{i}{2} \{ \Sigma^< \stackrel{\otimes}{,} S^>  \} (x,y)  \  , \end{align}
where $\Re$ refers to the real part, and
the first commutator is to be understood as 
\be [ S_{0}^{-1} - \Sigma^{\rm t} , S^<] (x,y) \equiv [S_{0,x}^{-1} - \Sigma^{\rm t} (x)] S^< (x,y) - S^<(x,y) [ (S_{0,y}^{-1})^\dag - \Sigma^{\rm t} (y) ] \ . \ee
In Eq.~(\ref{diffe-eq}) we have also introduced the notation,
\begin{align}  [A \stackrel{\otimes}{,} B] & \equiv   A\otimes B - B\otimes A \ , \\
\{ A \stackrel{\otimes}{,} B \} & \equiv   A \otimes B + B \otimes A \  ,\end{align}
and used the relation $\Re S^R  =\frac{1}{2} (S^R +S^A) $, and
$\Re \Sigma^R= \frac{1}{2} (\Sigma^R +\Sigma^A)  $.

Eq.~(\ref{diffe-eq}) is a generic equation for a theory of relativistic fermions, where all operators carry their own Dirac structure. For example, in QED one would use $S_{0,x}^{-1}=i \slashed{D}_{x}=i\slashed{\partial}_x-e\slashed{A}(x)$. Incidentally, the expression for the scalar case is totally analogous but the spin structure disappears~\cite{Blaizot:2001nr}. 

In QED the photon propagator also obeys a similar equation to (\ref{eq:KB1})~\cite{Blaizot:1999xk}
 \be \label{eq:photprop}
 \left( g^{\mu \nu} \partial^2 - \partial^\mu \partial^\nu - \Pi^{\rm t,\mu \nu} \right) D_{\nu \rho}^< (x,y) = \left ( \Pi^R  \otimes D^<\right)^\mu_\rho (x,y)  + \left(\Pi^<  \otimes D^A \right)^\mu_\rho (x,y) \ ,
 \ee
where $\Pi^{\mu \nu}$ is the photon (resp. tadpole, retarded and lesser) self-energy.

In order to derive the transport equations one first defines the (gauge covariantly modified)  Wigner function
\be \label{eq:Wigfun} S^<(X, K) = \int d^4 s \ e^{i K \cdot s} U \left(X, X+ \frac s2 \right) S^< \left( X+ \frac s2, X - \frac s2 \right) U \left(X- \frac s2, X \right)   \ , \ee
where $s=x-y$ and $X=(x+y)/2$ are the relative and center-of-mass coordinates, respectively, and $U(x,y)$ is the Wilson line joining $x$ and $y$.
The Wigner function is a function of $X^\mu$ and the  fermion kinetic momentum $K^\mu$.

Applying the (gauge covariantly modified) Wigner transform to the whole set of Kadanoff-Baym equations, and performing a gradient expansion, one ends up with a set of transport equations describing the system.

An important limitation of the above framework is that the transport approach is meant to 
describe the long distances, larger than the Compton wavelength for massive particles, or longer than the de Broglie wavelength for massless particles~\cite{relativisticbook,Blaizot:1999xk}, in order to avoid the effects of the ZB oscillations which occur at those scales.
 
In the remaining part of this manuscript we will use the same basic set of equations where we will treat the fermions using OSEFT, instead of the full QED theory.

\subsubsection{Comment on the dispersion relation}

 Before applying Eq.~(\ref{diffe-eq}) to the OSEFT, we briefly comment on the computation of the fermion dispersion relation, and argue that collisions do not modify it to the order we consider in this work.  On one hand the difference between Eqs.~(\ref{eq:KB1}) and~(\ref{eq:KB2}) gives the transport equation,  on the other hand their sum results in an independent equation, which in the collisionless case~\cite{Son:2012wh,Carignano:2018gqt} provided the fermion dispersion relation. While there is nothing wrong with this procedure, it is not the conventional way to address the dispersion relation. In fact, when collisions are present this is a more complicated path. The standard way is to consider the sum of the equation of motion of $S^R(x,y)$ and its conjugate and, after a Wigner transform, look for the poles of the retarded Green function. 

The equations of motion for $S^R(x,y)$ and its conjugate read (they can be obtained from the equations for $S^<(x,y)$ and $S^>(x,y)$, cf. Eq.~(\ref{eq:Sret}))
\begin{align} S_{0,x}^{-1}  \ S^R (x,y) & =\delta^{(4)}(x-y) + (\Sigma^R \otimes S^R )(x,y)   \ , \label{eq:GR} \\
  S^R (x,y) \ (S_{0,y}^{-1})^\dag & =  \delta^{(4)}(x-y) + ( S^R \otimes \Sigma^R)  (x,y) \ , \label{eq:GRconj}
\end{align}
where $\delta^{(4)}$ is the Dirac delta function. 
Notice that the terms which lead to collisions in the difference equation---which are those in the right hand side (RHS) of Eq.~(\ref{diffe-eq})---cancel. After half-summing~(\ref{eq:GR}) and~(\ref{eq:GRconj}) and performing the Wigner transform and gradient expansion (see later), one obtains the expression for $S^R (X,K)$. It is easy to check that the pole of $S^R(X,K)$ coincides with the solution of the constraint equation given by the sum equation of $S^<(x,y)$ and its conjugate---also after Wigner transform and gradient expansion---in the absence of collisions~\cite{Son:2012wh,Carignano:2018gqt}. The dispersion relation is the one found previously in Ref.~\cite{Carignano:2018gqt}. In that calculation, as well as in the present one, we will neglect the terms $\Sigma^{\rm t}$ and $ \Sigma^R$ as they provide contributions to the dispersion relation suppressed by   $\alpha$, the electromagnetic structure constant. It is also suppressed by $\sim \sqrt{\alpha}$ in comparison to the magnetic moment correction found in~\cite{Carignano:2018gqt}.

\section{Transport approach associated with the OSEFT~\label{sec:transpOSEFT}}

The formulation presented in the previous section is completely general, and therefore, it is also applicable to the OSEFT. However, we need to point out several particularities of this effective theory not present in the full QED theory. Therefore, let us recall some of the basic properties of the OSEFT in our original formulation. For more explicit details see
Ref.~\cite{Carignano:2018gqt}.

One first assumes that the full momentum $K^\mu$ of the fermion can be divided into a large part, and a residual or off-shell part $k^\mu$ as
\be \label{eq:momentum}
K^\mu =  E v^\mu + k^\mu \ , \qquad k ^\mu \ll E \ .
\ee

Let us recall how the Dirac and OSEFT fields are related. We express the Dirac field of a fermion with light-like velocity $v^\mu$,
\begin{equation}
\psi_{ v, \tilde v}=e^{-iE v\cdot x}\left( P_v \chi_{v}(x) +  P_{\tilde v} H_{\tilde v}^{(1)}(x)\right)+e^{i E\tilde{v}\cdot x} \left( P_{\tilde v} \xi_{\tilde v}(x)+P_v   H_{v}^{(2)}(x) \right)\ ,
\label{eq:Fields}
\end{equation}
where $P_v/P_{\tilde v}$ are particle/antiparticle projectors, respectively. The light-like velocity ${\tilde v}^\mu$ is fixed with the knowledge of the frame time-like vector
$u^\mu$, as  $u^\mu = (v^\mu + {\tilde v}^\mu)/2$. By integrating out the $H_{\tilde{v}}^{(1)}, H_v^{(2)}$ fields, we get a theory where the particle $\chi_v$ and antiparticle $\xi_{\tilde v}$ fields are totally decoupled. There is an interesting symmetry between the particle and antiparticle sectors of the theory, as all the equations for the antiparticles can be obtained from those of the particles after carrying out the changes $E, v^\mu \leftrightarrow -E, {\tilde v}^\mu$. For this reason we will concentrate our discussion on the particles, as all the equations for the antiparticles can be easily recovered.

We can still use the equations and formulation of Sec.~\ref{Sec-GeneralTransport}
if we understand that the two-point functions and fermion self-energy refer to the $\chi_v$, describing particles, or to $\xi_{\tilde v}$ if one wants to describe antiparticles. This encodes the whole discussion between Pauli-Dirac representation and the OSEFT (Foldy-Wouthuysen) one.
Thus, the starting two-point functions are
\be \label{eq:2pointOSEFT}
S_{E,v}^< (x,y) =  - \langle \bar \chi_v (y)  \chi_v(x) \rangle 
\ee
for particles, or
\be \label{eq:2pointOSEFT-anti}
{\tilde S}_{-E,{\tilde v}}^< (x,y) =  - \langle \bar \xi_{\tilde v} (y)  \xi_{\tilde v} (x) \rangle 
\ee
for antiparticles. In the Fourier space, the OSEFT fields are functions of the residual momentum, and thus also
the associated two-point functions.
 As said before, we will mainly focus on the particles in the following.

A main difference with the QED case is that instead of using as the inverse free propagator $S^{-1}_{0x}=i\slashed{D}_x$, the OSEFT Lagrangian indicates that this has
to be replaced by
\be S^{-1}_{0x} \rightarrow {\cal O}_x =  \left( i v\cdot D_x +\frac{(i\slashed{D}_{x,\perp})^2}{2E} + \cdots \right) \frac{\slashed{\tilde{v}}}{2} \ . \ee
written as an expansion in $1/E$. In particular, in Ref.~\cite{Carignano:2018gqt} we explicitly computed up to order $n=2$ in the energy expansion,
\be 
{\cal O}_x \ S^{<}_{E,v}(x,y)  -  S^{<}_{E,v} (x,y)  \ {\cal O}^\dag_y
\ee
and checked that after a Wigner transformation, it would give rise to the chiral transport equation.

At the end of the computations, and to compare with the full theory, we  re-express  all our final results in terms of the full momentum $K^\mu$. Actually, this last operation acts as
 a sort of consistency check of the computations, as the dependence on the residual momentum of all the final physical quantities should disappear.

In the remaining part of the manuscript we will see  how the collision terms of the CKT can be derived by considering the fermion self-energy corrections in the OSEFT.

\section{Collision terms and CKT from OSEFT~\label{sec:colloseft}}

In this section we proceed with the calculation of the collision terms of the OSEFT transport equation associated with the particle field. The collision term that enters into the antiparticle transport equation can be easily deduced from the particle one. 

In the OSEFT we start from the two-point function $S^<_{E,v} (x,y)$, and the Wigner function
\be \label{eq:Wigner}
S^<_{E,v} (X, k) = \int d^4 s \ e^{i k \cdot s} \ U \left(X, X+ \frac s2 \right) S^< \left( X+ \frac s2, X - \frac s2 \right) U \left(X- \frac s2, X \right) \ ,
\ee
is given in terms of the residual momentum (cf. Eq.~(\ref{eq:momentum})).

We then apply a Wigner transformation and a gradient expansion to Eq.~(\ref{diffe-eq}), taking also the Dirac trace. Let us first discuss the LHS of Eq.~(\ref{diffe-eq}). As argued before we neglect terms which are suppressed at weak coupling, like $\textrm{Re } \Sigma^R$
which would introduce $\alpha$ corrections to both the dispersion relations and LHS of the transport equation.
 The LHS is finally written as a $1/E$ expansion~(this calculation was performed up to order $1/E^2$ in Ref.~\cite{Carignano:2018gqt} to obtain the collisionless transport equation)
\be \label{eq:LHS_G} \left( 2iv^\mu  + 2 \frac{i}{E} k_{\perp}^\mu  + \cdots \right) \Delta^{k}_\mu G^\chi_{E,v} (X,k) \ , \ee
where 
\be
\Delta^{k}_\mu= \frac{\pa}{\pa X^\mu}-eF_{\mu\nu}(X)  \frac{\pa}{\pa k^\nu} \ 
\ee
is the transport operator.

To obtain (\ref{eq:LHS_G}) we recall that in the OSEFT the Wigner function can be expressed as
\be S^{<}_{E,v} (X,k) = \sum_{\chi =\pm } P_{\chi} \slashed{v} \, G^{<,\chi}_{E,v} (X,k) \ , \label{eq:S12} \ee
where we introduce the chirality projectors,
\be \label{eq:proj}
P_\chi = \frac{1 + \chi \gamma_5}{2} \ , \qquad \chi = \pm \ ,
\ee 
and
\be
\label{G-function}
G^{<,\chi}_{E,v} (X,k) = (2 \pi) \delta ({\cal K}^\chi_{E,v}) f^\chi_{E,v} (X, k) \ ,
\ee
written in terms of the distribution function $f^\chi_{E,v}(X,k)$.  We keep the labels of $E$ and $v$ in the distribution function, as in thermal equilibrium, for example,
the energy $E$ acts as a sort of  chemical potential for the modes with residual momenta $k$ \cite{Manuel:2016wqs}.
The Dirac delta function puts the particle on-shell, and the function  ${\cal K}^\chi_{E,v}$  fixes the dispersion relation to a certain order in the $1/E$ expansion~\cite{Carignano:2018gqt}. For example, at $n=1$
\be
 {\cal K}^\chi_{E,v} =  2 k \cdot v + \frac{1}{E}  \left(  k^2_{\perp} - \frac{e \chi}{4} \epsilon^{\alpha \beta \mu \nu}  {\tilde v}_\beta v_{\alpha} F^\perp_{\mu \nu} (X)
  \right) + {\cal O} \left( \frac{1}{E^2} \right) \ .
\ee

For later reference the two-point function $S^{>}_{E,v} (X,k)$ is expressed in the same manner, by replacing $ f^\chi_{E,v} (X, k)$ by $ 1 - f^\chi_{E,v} (X, k)$. 
 If we write explicitly Eq.~(\ref{eq:LHS_G}) up to ${\cal O}(1/E^2)$ in terms of the full momentum~(\ref{eq:momentum}), the fermion total energy
\be \label{eq:EK1} E_{K} = K \cdot u = E + k \cdot u \ , \ee
and the on-shell velocity $v^\mu_{K}$,
\be \label{eq:vK1} v^\mu_{K}=\frac{K^\mu}{E_{K}}
\ , \ee
we obtain
\be \label{eq:LHS} 2 i \left[ v_{K}^\mu - \frac{e}{2E_{K}^2} S^{\mu \nu}_{\chi} F_{\nu \rho} (X) (2u^\rho-v_{K}^\rho) \right] \Delta^{K}_{\mu} G^\chi(X, K) \ ,
\ee
where the spin tensor is
\be \label{eq:spin}S^{\mu \nu}_{\chi} \equiv \frac{\chi}{2} \epsilon^{\alpha \beta \mu \nu} \frac{u_\beta K_{\alpha}}{u \cdot K} = \frac{\chi}{2} \epsilon^{\alpha \beta \mu \nu} \frac{\tilde{v}_{\beta}}{2} v_{\alpha} + {\cal O} \left( \frac{1}{E} \right) \ , \ee
and we have converted
\be \label{eq:Gfull}
G^{<,\chi}_{E,v} (X,k)  =G^{<,\chi} (X,K) \ , \qquad G^{>,\chi}_{E,v} (X,k)   =G^{> \chi} (X,K) 
\ee
into functions of the full momenta~\cite{Carignano:2018gqt}.

Before introducing explicitly the distribution function let us look into the RHS of Eq.~(\ref{diffe-eq}). After a Wigner transform and Dirac trace we obtain
\be i \textrm{ Tr} \left[ \Sigma_{E,v}^> (X,k)  S_{E,v}^<(X,k) -  \Sigma_{E,v}^<(X,k)  S_{E,v}^>(X,k)   \right] \  , \ee
which will be computed up to order $1/E$.

Let us call this RHS the {\it total collision term}
\be iC^T = iC^{\rm gain} + iC^{\rm loss}  = i \textrm{Tr } [- \Sigma_{E,v}^< (X,k) S_{E,v}^> (X,k) + \Sigma_{E,v}^> (X,k) S_{E,v}^< (X,k) ] \ , \label{eq:initialtrace} \ee
where the first and second terms in the trace correspond to gain and loss terms, respectively.  As we concentrate in the transport equation for particles, this collision term will describe binary particle-particle and particle-antiparticle collisions. We focus here on the gain term of this equation, as the loss term is computed in an analogous way and will be added later too.

\begin{figure}[tp!]
\begin{center}
\includegraphics[scale=0.6]{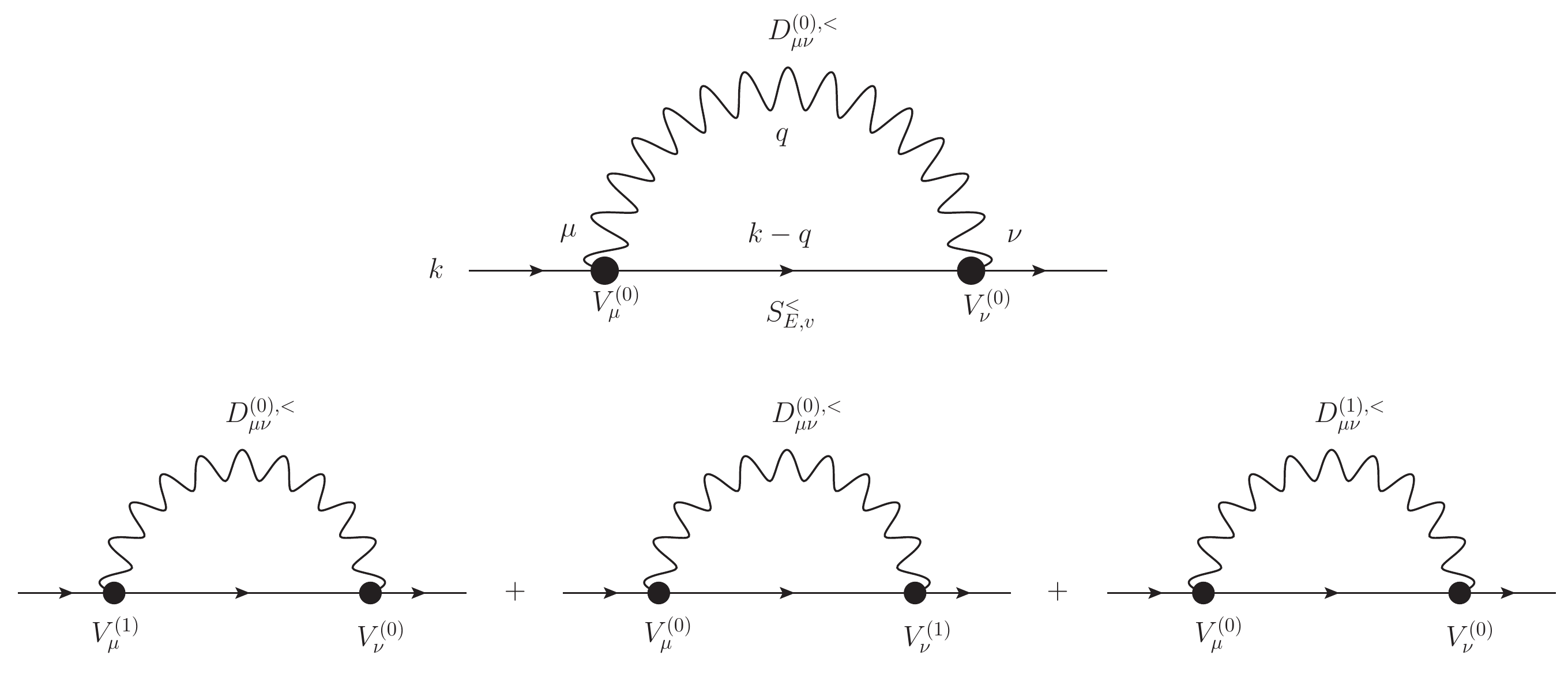}
\caption{Diagrammatic terms of the particle self-energy in~(\ref{eq:trace}), where the fermion vertices are denoted by their order $n=0$ or $n=1$. }
\label{fig:trace}
\end{center}
\end{figure}

At order $n=1$ the fermion self-energy contains four terms due to the possible combination of vertices and photon propagator at order $n=0$ and $n=1$ (see Fig.~\ref{fig:trace} for illustration). With this terms the gain part of the collision term is
\begin{align} 
C^{\rm gain} & = -i \int  \frac{d^4q}{(2\pi)^4}  \textrm{Tr } [V^{\mu,(0)} S_{E,v}^< (X,k-q) V^{\nu,(0)} S^{>}_{E,v} (X,k) ] D^{(0),<}_{\mu \nu} (X,q) \nonumber \\
& - i \int  \frac{d^4q}{(2\pi)^4}  \textrm{Tr } [V^{\mu,(1)} S_{E,v}^< (X,k-q) V^{\nu,(0)}   S^{>}_{E,v} (X,k) ] D^{(0),<}_{\mu \nu} (X,q) \nonumber \\
& -i  \int  \frac{d^4q}{(2\pi)^4} \textrm{Tr }[V^{\mu,(0)} S_{E,v}^< (X,k-q) V^{\nu,(1)}  S^{>}_{E,v} (X,k)] D^{(0),<}_{\mu \nu} (X,q) \nonumber \\
& - i  \int \frac{d^4q}{(2\pi)^4} \textrm{Tr }[ V^{\mu,(0)} S_{E,v}^< (X,k-q) V^{\nu,(0)}  S^{>}_{E,v} (X,k)]D^{(1),<}_{\mu \nu} (X,q) \ . \label{eq:trace}
\end{align}
where the vertices at $n=0$ and $n=1$ orders reads~\cite{Manuel:2016cit,Carignano:2018gqt} \footnote{While the Feynman rules in Ref.~\cite{Manuel:2016cit} were given in the frame associated at rest with the plasma, in Ref.~\cite{Carignano:2018gqt} it was explained how to generalize those to an arbitrary frame.}
\begin{align} \label{eq:vertices}
 V^{\mu,(0)} & = ie \frac{\slashed{\tilde{v}}}{2} v^\mu \ , \\ 
 V^{\mu,(1)} & = i\frac{e}{E} \frac{\slashed{\tilde{v}}}{2} \left[ (k_\perp^\mu + \frac{1}{2} q_\perp^\mu) - \frac{i}{2} \sigma_\perp^{\mu \alpha}q_\alpha\right] \ ,
 \end{align}
where $q^\mu$ is the momentum of the incoming photon, and $k^\mu$ the residual momentum of the incoming fermion. 

Let us mention that the two point-functions $S_{E,v}^<$ and $S_{E,v}^>$ should also be expanded in $1/E$ by expanding the Dirac delta function of Eq.~(\ref{G-function}). As in
Ref.~\cite{Manuel:2016wqs}, we keep the general structure of these functions without expanding the delta function, as this facilitates both the intermediate computations and 
the expressions of the final results in terms of the full momentum, but  one should keep in mind this fact when doing the power counting in $1/E$.

The first line of (\ref{eq:trace}) (upper diagram of Fig.~\ref{fig:trace}) contains the LO ($n=0$) pieces for the vertices and photon propagators, while the three last terms of that equation (lower diagram in Fig.~\ref{fig:trace}) represent the NLO ($n=1$) contributions of vertices and photon propagators to the trace. 

 Eq.~(\ref{eq:trace}) is written in terms of a  photon two-point function. This can be obtained after carrying out a Wigner transformation and a gradient expansion to 
Eq.~(\ref{eq:photprop}) so that one ends up with  \cite{Blaizot:1999xk} 
 \be \label{eq:2pointphoBIt} D^{<}_{\mu \nu} (X,q) \simeq D^R_{\mu \rho} (X,q) \ \Pi^{<,\rho \sigma} (X,q) \ D^A_{\sigma \nu} (X,q) \ ,  \ee
in terms of the retarded and advanced photon propagators.
 We use this result by noting that both the photon polarization, and thus also the photon propagator, can be computed in OSEFT in a $1/E$ expansion.

 At order $n$ the photon polarization tensor  reads
\ba \label{eq:photonpol} \Pi^{(n),<}_{\rho \sigma} (X,q) &= &i  \sum_{j=0 }^n \sum_{E',v'}  \int \frac{d^4k'}{(2\pi)^4} \textrm{Tr } [ V^{(j)}_\rho  S_{E',v'}^< (X,q+k') \ V_\sigma^{(n-j)} \  S^{>}_{E',v'} (X,k')]  \nn
\\
  &+ &i \sum_{j=0}^n \sum_{E',{\tilde v}'}  \int \frac{d^4k'}{(2\pi)^4} \textrm{Tr } [{\tilde V}^{(j)}_\rho  {\tilde S}_{-E',{\tilde v'}}^< (X,q+k') \ {\tilde V}_\sigma^{(n-j)} \  {\tilde S}^{>}_{-E',{\tilde v}'} (X,k')] \ , 
 \ea
where the first/second integral arises from  particle/antiparticle contributions to the photon polarization tensor. 

The first sum in the above integrals counts the possible combination of vertices leading to desired order. The second sum is performed together with the integration of the residual momentum $k'$. It is worth reminding here that $q$ is the (soft) photon momentum, and not a residual momentum.

\begin{figure}[tp!]
\begin{center}
\includegraphics[scale=0.6]{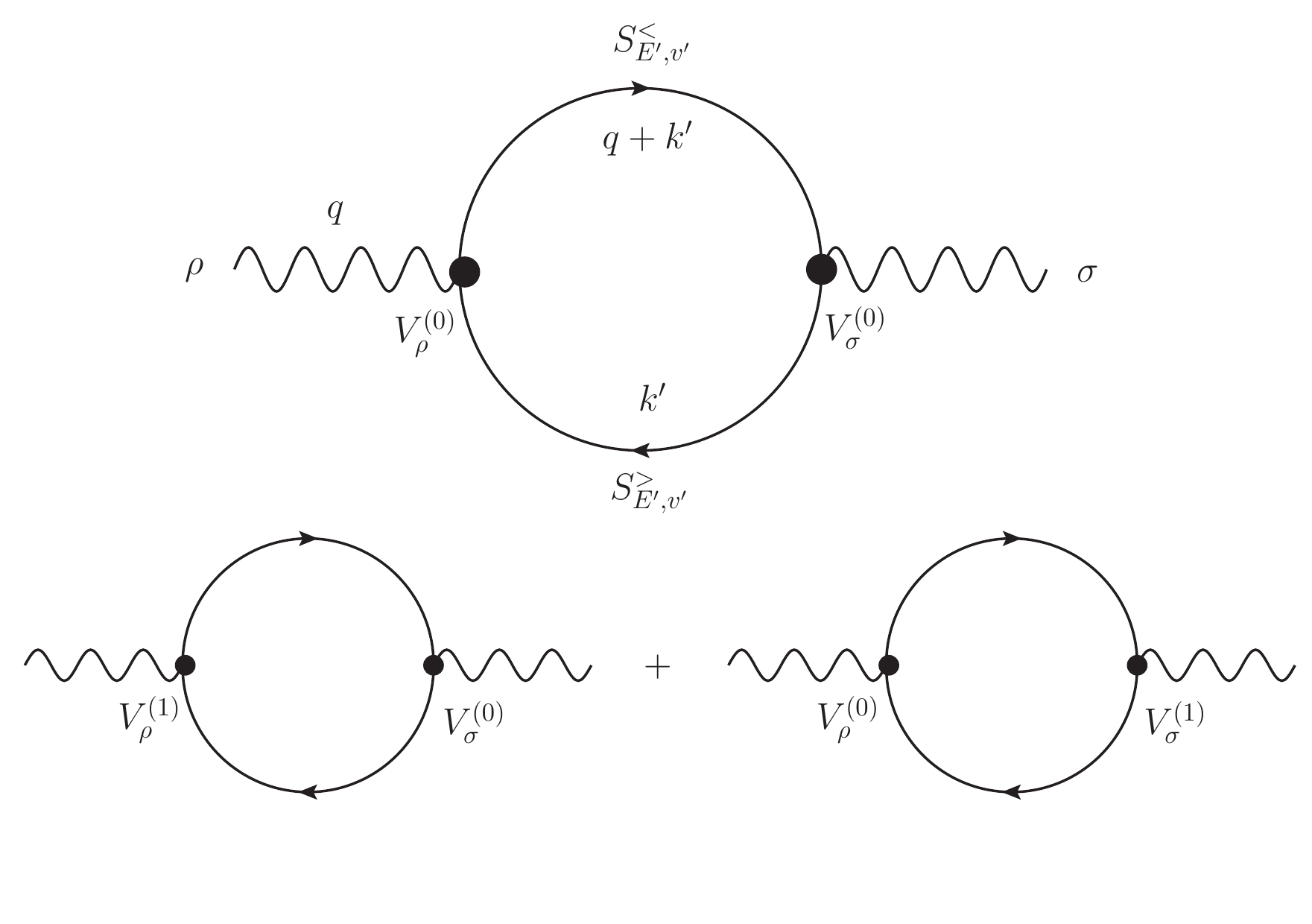}
\caption{Photon polarization function up to $n=1$, where 3 diagrams contribute. The fermion vertices are denoted by their order $n=0$ or $n=1$. }
\label{fig:polfuncs}
\end{center}
\end{figure}

For $n=0$ only one diagram contributes to the photon polarization function, whereas for $n=1$ two diagrams must be considered. We plot them in Figure~\ref{fig:polfuncs}.

Let us give some details of the technical calculation, where we will only focus on the particle contribution, and only add the antiparticle contribution at the very end. Let us start computing the photon polarization function at $n=0$. There is a single diagram, constructed by two $n=0$ vertices,
\be \Pi^{(0),<}_{\rho \sigma} (X,q) = - i e^2 \sum_{E',v'} \int \frac{d^4k'}{(2\pi)^4} v'_\rho v'_\sigma \ \textrm{Tr} [\frac{\slashed{\tilde{v}'}}{2} S_{E',v'}^< (X,q+k') \frac{\slashed{\tilde{v}'}}{2} S^{>}_{E',v'} (X,k')] \ . \ee

After carrying out the Dirac trace one realizes that different chiralities are not mixed up inside the loop, as the  photon cannot produce a chirality flip in the vertex. Thus, one finds
\begin{align}
   \Pi^{(0),<}_{\rho \sigma} (X,q) & = - 4i e^2 \sum_{E',v'} \int \frac{d^4k'}{(2\pi)^4} 
   v'_\rho v'_\sigma \sum_{\chi'=\pm}   \ G_{E',v'}^{<,\chi'} (X,q+k') G_{E',v'}^{>,\chi'} (X,k') \ .
\end{align}

Using this result, and the one for the case $n=1$ we can obtain the photon propagator at LO and NLO 
\begin{align}
 D^{(0),<}_{\mu \nu} (X,q) & = -4ie^2 D^R_{\mu \rho} (X,q) D^A_{\sigma \nu} (X,q) \sum_{E',v'}  \int \frac{d^4k'}{(2\pi)^4} \sum_{\chi'=\pm} v'^\rho v'^\sigma G_{E',v'}^{<,\chi'} (X,q+k') G_{E',v'}^{>,\chi'} (X,k') \ , \label{eq:photonLO}\\ 
  D^{(1),<}_{\mu \nu} (X,q) & = -ie^2 D^R_{\mu \rho} (X,q) D^A_{\sigma \nu} (X,q) \nonumber \\ 
&\times  \sum_{E',v'}  \int \frac{d^4k'}{(2\pi)^4} \sum_{\chi'=\pm }  \frac{1}{E'}  \left\{ 
  4 v'^\sigma  (k_\perp'^{\rho} + \frac{1}{2} q_\perp^\rho) -  \chi' v'^\sigma  q_{\perp,\alpha} i \epsilon^{\lambda \omega \rho_\perp \alpha} \tilde{v}'_\lambda   v'_\omega  \right. \nonumber \\
&\left. +
 4 v'^\rho(k_\perp'^{\sigma} + \frac{1}{2} q_\perp^\sigma) +  \chi' v'^\rho q_{\perp,\alpha} i \epsilon^{\lambda \omega \sigma_\perp \alpha} \tilde{v}'_\lambda   v'_\omega
\right\}
 G_{E',v'}^{<,\chi'} (X,q+k') G^{>,\chi'}_{E',v'} (X,k') \ ,  \label{eq:photonNLO}
\end{align}
respectively, where in the latter case some terms depend explicitly on the chirality $\chi'$, but it is still conserved in the vertex.

The only missing step is to combine all the computed pieces, and express the result in a familiar way, in terms of the fermion distribution function, expressed in terms of the full momentum, instead of the residual momentum used in the OSEFT. We denote with capital/lowercase letters the full/residual momenta, respectively. We make an exception with the photon momenta, which is denoted with lowercase letters because it is soft and of the same order than the residual momenta.
Recall that from Eq.~(\ref{eq:vK1}) the on-shell velocity differs from $v^\mu$ already at order $1/E$.
 
The soft photon propagator at LO+NLO is the combination of (\ref{eq:photonLO}) and (\ref{eq:photonNLO}).
\begin{align}
 D^{(0),<}_{\mu \nu} (X,q) & + D^{(1),<}_{\mu \nu} (X,q)  = -2ie^2 D^R_{\mu \rho} (X,q) D^A_{\sigma \nu} (X,q)   \nonumber \\ 
&\times  \sum_{E',v'} \int \frac{d^4k'}{(2\pi)^4} \sum_{\chi'=\pm} \left\{ 
   v'^{\sigma}_{K'} 
  \left(v'^{\rho}_{K'}  + \frac{q_\perp^\rho}{E_{K'}} \right)  +  
  v'^{\rho} _{K'}   \left(v'^{\sigma}_{K'}  + \frac{q_\perp^\sigma}{E_{K'}} \right) \right. \nonumber \\
&\left. -  \frac{i}{2E_{K'}} \chi' v'^{\sigma}_{K'}  q_{\perp,\alpha} \epsilon^{\lambda \omega \rho_\perp \alpha} \tilde{v}'_\lambda   v'_\omega
  + \frac{i}{2E_{K'}} \chi' v'^{\rho}_{K'} q_{\perp,\alpha} \epsilon^{\lambda \omega \sigma_\perp \alpha} \tilde{v}'_\lambda   v'_\omega    
\right\} \nonumber \\
&\times G_{E',v'}^{<,\chi'} (X,q+k') G^{>,\chi'}_{E',v'} (X,k') \ ,
\end{align}
where we have combined already some pieces in terms of full momenta $K'^{\mu}$.

One can insert the photon propagator into each of the four traces of Eq.~(\ref{eq:trace}). After performing the Dirac traces that appear in that equation, and half dozen of steps to combine and simplify terms we arrive to the result
\begin{align}
C^{\rm gain} & = 4e^4 \sum_{\chi=\pm} \int \frac{d^4q}{(2\pi)^4} \sum_{E',v'} \int \frac{d^4k'}{(2\pi)^4}   D^R_{\mu \rho} (X,q) D^A_{\sigma \nu} (X,q)    \nonumber \\ 
& \times \sum_{\chi'=\pm } \left\{ 
v_{K}^\mu \left( v_{K}^\nu-\frac{q^\nu_\perp}{E_{K}} \right) +v_{K}^\nu \left( v_{K}^\mu-\frac{q^\mu_\perp}{E_{K}} \right)  
- \frac{2i}{E_{K}} v_{K}^\mu q_{\perp,\alpha} S_\chi^{\alpha \nu_\perp}   + \frac{2i}{E_{K}} v_{K}^\nu q_{\perp,\alpha}   S_\chi^{\alpha \mu_\perp}  
\right\} \nonumber \\ 
& \times \left\{ 
   v_{K'}^\sigma   \left(v_{K'}^\rho  + \frac{q_\perp^\rho}{E_{K'}} \right)  +  
  v_{K'}^\rho   \left(v_{K'}^\sigma  + \frac{q_\perp^\sigma}{E_{K'}} \right)  
  -  \frac{2i}{E_{K'}}  v_{K'}^\sigma   q_{\perp,\alpha}  S_{\chi'}^{\alpha \rho_\perp}   + \frac{2i}{E_{K'}} v_{K'}^\rho q_{\perp,\alpha}  S_{\chi'}^{\alpha \sigma_\perp} \right\} \nonumber \\
& \times G_{E,v}^{<,\chi} (X,k-q)  G^{>,\chi}_{E,v} (X,k) G_{E',v'}^{<,\chi'} (X,q+k') G^{>,\chi'}_{E',v'} (X,k')  \ ,
\end{align}
where we have introduced the spin tensor~(\ref{eq:spin}). Notice that the Dirac trace in~(\ref{eq:trace}) gives a sum over the chiralities $\chi$ of the fermion with momentum $K$. As done in Ref.~\cite{Carignano:2018gqt} we simply focus on a fermion with some particular chirality $\chi$, and write the collision term for this particular particle.  We stress that this chirality is not modified in the interaction with a soft photon.

To further simplify this expression we define the scattering amplitude squared, which is already a function of the full momenta and the photon soft momentum as
\begin{align}
&|M_{\chi,\chi'}|^2 (K,K',q) \equiv 4e^4  E_{K} E_{K-q} E_{K'} E_{q+K'} D^R_{\mu \rho} (X,q) D^A_{\sigma \nu} (X,q)   \nonumber \\ 
& \times  \left\{ 
v_{K}^\mu \left( v_{K}^\nu-\frac{q^\nu_\perp}{E_{K}} \right) +v_{K}^\nu \left( v_{K}^\mu-\frac{q^\mu_\perp}{E_{K}} \right)  
- \frac{2i}{E_{K}} v_{K}^\mu q_{\perp,\alpha} S_\chi^{\alpha \nu_\perp}   + \frac{2i}{E_{K}} v_{K}^\nu q_{\perp,\alpha}   S_\chi^{\alpha \mu_\perp}  
\right\} \nonumber \\ 
& \times \left\{ 
   v_{K'}^\sigma   \left(v_{K'}^\rho  + \frac{q_\perp^\rho}{E_{K'}} \right)  +  
  v_{K'}^\rho   \left(v_{K'}^\sigma  + \frac{q_\perp^\sigma}{E_{K'}} \right)  
  -  \frac{2i}{E_{K'}}  v_{K'}^\sigma   q_{\perp,\alpha}  S_{\chi'}^{\alpha \rho_\perp}   + \frac{2i}{E_{K'}} v_{K'}^\rho q_{\perp,\alpha} S_{\chi'}^{\alpha \sigma_\perp} \right\} \ . \label{eq:Msq}
\end{align}
Here we have explicitly included the factors $E_{K} E_{K-q} E_{K'} E_{q+K'}$ in order to have an adimensional scattering amplitude squared, following standard relativistic quantum field theory conventions. These precise energy factors will later allow us to eventually express the collision term, when given on shell, in terms of Lorentz-invariant phase space measures.
  
In terms of this amplitude, and when complemented with the loss term, the total collision term of the particle-particle interaction is
\begin{align} 
C^T &=  \int \frac{d^4 q}{(2\pi)^4} \int \frac{d^4 K'}{(2\pi)^4} \sum_{\chi'=\pm} |M_{\chi,\chi'}|^2 (K,K',q) \ \frac{1}{E_{K}  E_{K-q} E_{K'} E_{q+K'}} \nonumber \\ 
& \times  \left[ G^{>,\chi} (X,K) G^{<,\chi} (X,K-q) G^{<,\chi'} (X,q+K') G^{>,\chi'} (X,K') \right. \nonumber  \\
&- \left. G^{<,\chi} (X,K) G^{>,\chi} (X,K-q) G^{>,\chi'} (X,q+K') G^{<,\chi'} (X,K') \right] \ .  
\end{align}
where we have used the prescription~\cite{Manuel:2016wqs}
\be \sum_{E',v'} \int \frac{d^4 k'}{(2\pi)^4} = \int \frac{d^4K'}{(2\pi)^4} \ , \ee
and we have converted all $G^{<,\chi}_{E,v} (X,k)$ into $G^{<,\chi} (X,K)$~(\ref{eq:Gfull}) in terms of the full momentum.

At this point we note that by including the contributions to the photon polarization tensor as arising from the antiparticles, we can construct also the collision term that describes particle-antiparticle scatterings.

Combining the LHS of the kinetic equation~(\ref{eq:LHS}) and the RHS (equal to $iC^T$), the transport equation in an arbitrary frame reads
\begin{align} \label{eq:anyframe}
2 \left[ v_{K}^\mu - \frac{e}{2E_{K}^2} S^{\mu \nu}_{\chi} F_{\nu \rho} (X) (2u^\rho-v_{K}^\rho) \right] \Delta^{K}_{\mu} G^\chi(X, K) =  C^T +\tilde{C}^T \ ,
\end{align} 
where we added the particle-antiparticle collision term $\tilde{C}^T$, which is equal to the previous one but the Green functions involved in $\tilde{C}^T$ describes a collision of a particle degree of freedom with an antiparticle one.

We point out that at order $n=1$ the collision term depends on the spin tensor associated with the particle. One should also further notice that the photon propagator in a
medium with fermion chiral imbalance has three independent components, one associated with the longitudinal mode and two transverse different components, which correspond
to right- and left-handed circular polarizations (see App.~\ref{app-photon} for  explicit expressions close to equilibrium.) Let us mention that the scattering matrix element
at leading order is the same that one would expect in the Born approximation in QED for soft momentum transfers (see Eq.(2.3)  of Ref.~\cite{Blaizot:1996az}), while the new corrections 
proportional to the spin tensor will describe how a fermion of a given chirality interacts in a different way with the transverse photons of different helicity, as it will be explicitly seen in Sec.~\ref{sec:decayrate}.

\subsection{Collision term in the local rest frame}

It is possible to simplify the form of the collision term if one chooses the local rest frame with the plasma, $u^\mu = (1,0)$. In addition we need to express it in terms of the fermion distribution functions. The OSEFT Green functions, which are functions of the residual momentum, can also be  written in terms of the full momentum  as
\begin{align} 
G^{<,\chi}_{E,v} (X,k) & =G^{<,\chi} (X,K) = \pi \theta(E_K) \delta(K^0-E_K) f^\chi(X,K) \ , \\
G^{>,\chi}_{E,v} (X,k) & = G^{> \chi} (X,K)  =\pi \theta(E_K) \delta(K^0-E_K) [1-f^\chi(X,K)] \ ,
\end{align}
at order $n=1$.

The collision terms are now functionals of the distribution functions (which we will indicate explicitly as arguments of $C^T$), and they are written as functions of the full momenta,
\begin{align} 
C^T [f^\chi,f^{\chi'}] &=  \int \frac{d^4 K_2}{(2\pi)^3}  \frac{d^4 K_3}{(2\pi)^3} \frac{d^4 K_4}{(2\pi)^3} \sum_{\chi'=\pm} |M_{\chi,\chi'}|^2(K,K_2,q) \ \frac{2\pi}{2E_{K}  2E_{K_2} 2E_{K_3} 2E_{K_4} } \nn \\ 
& \times (2\pi)^4 \delta^{(3)} ({\bf K}+{\bf K}_2 - {\bf K}_{3} - {\bf K}_{4}) 
\delta (K^0+K^0_2 -  K^0_{3} - K^0_{4})  \nonumber \\ 
& \times  \left\{  \theta(E_{K}) \delta(K^0-E_{K}) (1-f^{\chi} (X,K) ) 
\ \theta(E_{K_3}) \delta(K_3^0-E_{K_3}) f^{\chi} (X,K_3) \right. \nn \\
& \left.  \times  \theta(E_{K_4}) \delta(K_4^0-E_{K_4}) f^{\chi'} (X,K_4) 
\ \theta(E_{K_2}) \delta(K_2^0-E_{K_2}) (1-f^{\chi'} (X,K_2))  \right. \nn  \\
&- \left.  \theta(E_{K}) \delta(K^0-E_{K}) f^{\chi} (X,K) 
\ \theta(E_{K_3}) \delta(K_3^0-E_{K_3}) f^{\chi} (X,K_3) \right. \nn \\
& \times  \left. \theta(E_{K_4}) \delta(K_4^0-E_{K_4}) (1-f^{\chi'} (X,K_4)) 
\ \theta(E_{K_2}) \delta(K_2^0-E_{K_2}) f^{\chi'} (X,K_2) 
\right\}   \ , 
\end{align}
where we have changed notation $K' \rightarrow K_2$, performed the change of variables $K_4 \equiv q +K_2$, and introduced $K_3$ via the identity:
\be 1=(2\pi)^4 \int \frac{d^4 K_3 }{ (2\pi)^4} \delta^{(3)} ({\bf K}+{\bf K}_2 - {\bf K}_{3} - {\bf K}_{4}) 
\delta (K^0+K^0_2 -  K^0_{3} - K^0_{4}) \ . \label{eq:unit} \ee

We can now integrate over all zero components of momenta $dK^0_2,dK^0_3,dK^0_4$ as well as over $dK^0/2\pi$. The result is a collision term for the {\it on-shell distribution function}.
\begin{align}
{\cal C}^T[f^\chi,f^{\chi'}] \equiv \int \frac{dK^0}{2\pi}  C^T [f^\chi,f^{\chi'}] 
\end{align}
where all energies are on shell and are functions of their respective momenta and magnetic field, and we have defined
\be
 \ippe{K_i} \equiv \int \frac{d^3 K_i }{ (2\pi)^3 2E_{K_i}} \ .
\ee
 
The only remaining steps are to express the LHS Eq.~(\ref{eq:anyframe}) as function of the distribution function in the local rest frame  and integrate over $dK^0/2\pi$. We obtain,
recalling that the electric and magnetic fields $E^i, B^i$ are related to the field tensor as
$F^{i0} = E^i$, $F^{ij} = - \epsilon^{ijk} B^k$,
\be
\label{staticCKT}
\left ( \Delta^{K}_0 + \hat K^i \left (1 + e \chi \frac{ {\bf B} \cdot {\bf \hat K}}{2 K^2} \right) \Delta^{K}_i  + e \chi \frac{\epsilon^{ijk} E^j \hat{K}^k - B^i_{\perp,{\bf K}}}{4 K^2}  \Delta_i
\right) f^\chi(X, {\bf K}) ={\cal C}^T[f^\chi,f^{\chi'}]+\tilde{\cal C}^T[f^\chi,\tilde{f}^{\chi'}]   \ ,
\ee
where $K=|{\bf K}|, {\bf \hat K}={\bf K}/K$, ${\bf B}_{\perp,{\bf K}}={\bf B}-{\bf \hat K} ({\bf B} \cdot {\bf \hat K})$, and
\begin{align} 
{\cal C}^T[f^\chi,f^{\chi'}] = & \frac{1}{2E_{K}} \ippe{K_2, K_3, K_4}
   \sum_{\chi'=\pm } |M_{\chi,\chi'}|^2 (K,K_2,q) \ (2\pi)^4 \delta^{(4)} (K+K_2 - K_3 - K_4)  \nonumber \\
& \times  \left\{ f^{\chi} (X,{\bf K}_3) f^{\chi'} (X,{\bf K}_4) [1-f^{\chi}(X,{\bf K})]  [1-f^{\chi'} (X,{\bf K}_2)] \right. \nonumber  \\
& - \left. f^{\chi} (X,{\bf K}) f^{\chi'}  (X,{\bf K}_2) [1-f^{\chi} (X,{\bf K}_3)] [1-f^{\chi'} (X,{\bf K}_4)] \right\} \ , \end{align}
with $|M_{\chi,\chi'}|^2 (K,K_2,q)$ given in Eq.~(\ref{eq:Msq}).

\section{Fermion decay rate in a chiral plasma~\label{sec:decayrate}}

In this section we review the computation of the decay rate of Ref.~\cite{Carignano:2018thu}, and check that the same result, at certain order of accuracy, can be obtained from
the collision term of the OSEFT just derived. For simplicity we will assume  a fermionic  system with chiral imbalance
 at $T=0$, with thus different right and left handed chemical potentials.
 (We make the assignments $\chi =+ = R$, and $\chi= - = L$).  We work in the local rest frame of the plasma. In this case the fermion distribution function becomes a step function
\be \label{eq:eqdistro}
f^\chi({\bf K}) = n^\chi_F (E_K) = \theta( \mu^\chi -E_{K}) \ .
\ee

We  compute the decay rate associated with  a  massless fermion with energy $E_{K}$, which is above the two Fermi surfaces, and with four momentum $K^\mu$, and chirality $\chi$. These systems suffer from the presence of chiral instabilities \cite{Akamatsu:2013pjd}. The computation of the fermion decay rate is only strictly valid for time scales much shorter than
the time scale of the onset of the instability $t_{\rm ins} \sim 1/\alpha^2 \mu_5$, where $\mu_5= \mu_+-\mu_-$  is the chiral chemical potential.
The decay rate can be computed from the imaginary part of the fermion self-energy, as done in Ref.~\cite{Carignano:2018thu}. The same computation can be performed
from the  kinetic theory of QED as
\be \label{eq:decayrate}
\Gamma^\chi (E_{K}) = \frac{(2\pi)^4 }{2E_{K}} \sum_{\chi' = \pm} \ippe{K_2,K_3,K_4}
 n^\chi_F(E_{K_2}) [ 1- n^{\chi'}_F(E_{K_4})] [1- n^\chi_F( E_{K_3})] \delta^{(4)} (K +K_2 - K_3  - K_4 ) |M^{\rm QED}_{\chi,\chi'}|^2
\ee
where we adopted the same variables as in previous section ($K,2 \rightarrow 3,4$).

The square of the scattering matrix of the process can be computed from QED, and it is given by
\be
\label{scatteringQED}
|M^{\rm QED}_{\chi,\chi'}|^2 = e^4 {\rm Tr}[\slashed{K_3}\gamma^\mu \slashed{K} \gamma^\rho {\cal P}_\chi ] D^R_{\mu\nu}(q) {\rm Tr}[\slashed{K_4}\gamma^\nu \slashed{K_2} \gamma^\lambda {\cal P}_{\chi'} ] D^A_{\rho\lambda}(q) 
\ee
where we have defined the momentum transfer by $q^\mu = (\omega, \qq)$. It is possible the eliminate the $\K{4}$ integral, using the conservation of momentum, and introduce
\be 
1 = \int d\omega \, \delta(\omega - E_{K_2+q} + E_{ K_2 } ) \int d^3q \, \delta^{(3)} (\qq - \K{} + \K{3} ) 
\ee
to trivially integrate integrate over $\K{3}$ and finally write
\ba
\Gamma^\chi (E_K) &=& \frac{\pi}{E_{K}} \sum_{\chi' = \pm}  \ippe{K_2} \ipp{q} \int d\omega \,\delta(\omega  - E_{ K_2+q} + E_{ K_2 } ) \delta(E_{ K } - E_{ K-q} - \omega) \nn
\\
&\times & \frac{n^{\chi'}_F ( E_{K_2}) [ 1- n^{\chi'}_F ( E_{K_2+q})] [1- n^\chi_F(  E_{K-q})]}{2 E_{K-q} 2 E_{K_2+q} } |M^{\rm QED}_{\chi,\chi'}|^2 
\ea

The Pauli blocking constraints of the problem impose that the momentum transfer has to be much smaller than the fermion energy and also than the two Fermi energies. We can then approximate
\be
n^{\chi'}_F( E_{K_2}) [ 1- n^{\chi'}_F( E_{K_2+q})]  = 
  \omega \delta(E_{K_2} - \mu^{\chi'}) + {\cal{O}}\Big(\Big(\frac{\omega}{E_{K_2}}\Big)^2\Big) 
\ee
which is already a $\omega/E_{K_{2}}$ effect.
 Eq.~(\ref{scatteringQED}) can also be worked out, and expanded in powers of both $1/E_{K}$ and $1/E_{K_{2}}$
\be
\label{ExpQEDscatt}
|M^{\rm QED}_{\chi,\chi'}|^2  =  e^4 D^R_{\mu\nu}(q)  D^A_{\rho\lambda}(q) 4 E_{K}^2 E_{K_2}^2 \Big[ 2 v_{K}^\mu v_{K}^\rho - v_{K}^\mu \frac{q^\rho}{E_{K}} - v_{K}^\rho \frac{q^\mu}{E_{K}} + i \chi \epsilon^{\alpha\mu\beta\rho} \frac{q_\alpha}{E_{K}} v_{K,\beta}\Big] 2 v_{K_2}^\nu v_{K_2}^\lambda + \cdots  \ ,
\ee
where we have introduced the notation $v_{K}^\mu = K^\mu/E_{K} = (1 , {\bf v}_{K})$, and  $v_{K_2}^\mu = K_2^\mu/E_{K_{2}} = (1 , {\bf v}_{K_2})$.
Notice that we neglect terms in the squared of the scattering matrix that depend on $\chi'$, as they would only contribute to order $1/E^2_{K_2}$ in the decay rate.
We further expand 
\be
E_{K-q} = E_{K} - \qpar + \frac{\qperp^2}{2E_{K}} + \dots  \ , \qquad  E_{K_2+q} = E_{K_2} - {\bf v}_{K_2} \cdot \qq + \dots 
\ee
where we defined $\qpar = {\bf v}_{K} \cdot \qq$, and ${\bf \qperp} = \qq - \qpar {\bf v}_{K} $.

The $|K_2|$ integration can be easily done
\be
\sum_{\chi' = \pm} e^2 \int \frac{ |K_2|^2 d |K_2|} {2 \pi^2}  \delta(E_{K_2} - \mu^{\chi'})=  m^2_D \ , 
\ee
 written in terms of the Debye mass, $m^2_D=e^2(\mu^2_R+\mu^2_L)/2\pi^2$. 
For the angular integrals of $\K{2}$ we use
\begin{align}
\label{angularint}
\int \frac{d\Omega_{2}}{4\pi} \delta(\omega - {\bf v}_{K_2} \cdot \qq)  & = \frac{1}{2|\qq|} \theta(\qq^2 - \omega^2) \ , \nn\\
  \int \frac{d\Omega_{2}}{4\pi} \delta(\omega - {\bf v}_{K_2} \cdot \qq)  v_{K_2}^i  & = \frac{1}{2|\qq|} \theta(\qq^2 - \omega^2) \frac{\omega}{|\qq|} {\hat q}^i  \ ,\nn\\
  \int \frac{d\Omega_{2}}{4\pi} \delta(\omega - {\bf v}_{K_2} \cdot \qq)  v_{K_2}^i v_{K_2}^j & = \frac{1}{2|\qq|} \theta(\qq^2 - \omega^2) \Big[ \frac{\qq^2 - \omega^2}{2|\qq|^2} \delta^{ij} + \frac{3\omega^2-\qq^2}{2|\qq|^2} {\hat q}^i{\hat q}^j \Big]  \ ,
   \end{align}
where $\Omega_2$ is the solid angle of the vector ${\bf v}_2$.
Using the explicit form of the retarded photon propagator (see Appendix~\ref{app-photon})
\be
\label{photon-propag}
D^R_{\mu\nu}(q) = \delta_{\mu 0}\delta_{\nu 0} {\cal D}_L(q) + \sum_{h=\pm} P^{T,h}_{ij}  {\cal D}_T^h(q) \delta_{\mu i}\delta_{\nu j}
\ee
where $h$ labels the two circular polarized transverse states, left and right, and we introduced  the transverse projector for a given helicity 
\be \label{eq:heliproj}
P_{ij}^{T,h} = \frac{1}{2} \Big( \delta_{ij} - {\hat q}_i{\hat q}_j - i h \epsilon_{ijk} {\hat q}_k \Big) \ .
\ee

The rate is  better  expressed in terms of the spectral functions of the longitudinal and transverse photon propagators on the chiral medium. Using \footnote{We write the
transverse spectral function as proportional to $m_D^2$ rather than  $M_h^2 = m_D^2 - he^2 \mu_5|\qq|/\pi^2$, as the difference can be amounted to the fact that we kept only the leading order in the 
$|\qq|/E_K$ expansion before doing the radial integral---in other words, the difference is a higher order effect we need not care for here.}
  
 \begin{align}
  \rho_L(\qpar,\qq) & =\pi m_D^2 \frac{\qpar}{|\qq|} \big\vert {\cal D}_L(\qpar, |\qq| ) \big\vert^2 \theta(\qq^2 - \qpar^2) \,, \\
  \rho_T^h(\qpar,\qq) & \approx \pi m_D^2 \frac{\qpar}{2|\qq|}  \Big(1 - \frac{\qpar^2}{|\qq|^2} \Big) \big\vert {\cal D}_T^h(\qpar, \qq ) \big\vert^2 \theta(\qq^2 - \qpar^2)  \ .
 \end{align}
  
So we can write 
\begin{align}
\label{interate} 
\Gamma^\chi (E_{K}) & =  \ipp{q}  \Big[ 1- n^\chi_F\big(E_{K} - \qpar \big)\Big] \nn \\ 
& \times \left\{  \Big( 1- \frac{\qpar}{E_{K}} \Big) \rho_L(\qpar,\qq)   + \sum_h
   \Big(1 - \frac{\qpar^2}{\qq^2} \Big) \Big[1 - \frac{1}{E_{K}} \big( \qpar + |\qq| \chi h \big) \Big]   \rho_T^h(\qpar,\qq) \right\} \ .
\end{align}

We thus reproduce the value of the damping rate of 
Eq.~(23) of Ref.~\cite{Carignano:2018thu}, considering that our $\Gamma = 2\gamma$ computed there. As in Ref.~\cite{Carignano:2018thu}, we see that a fermion of a given chirality interacts differently with transverse photons, depending on their helicity. This is an effect that is only non-vanishing in a medium where the transverse photon propagators of the two circular polarizations are different, as it is the case for a system with chiral fermion imbalance \cite{Nieves:1988qz}.

Within OSEFT we can derive the previous result, which we recall, has been computed to order $1/E$. The damping rate in OSEFT expressed in terms of the full momenta reads
\begin{align} 
\Gamma^\chi (E_{K}) & =  \frac{1}{2E_{K}} \ippe{K_2, K_3, K_4}
\sum_{\chi' =\pm } |M_{\chi,\chi'}|^2 (K,K_2,q) (2\pi)^4 \delta^{(4)} (K+K_2 -K_3 - K_4)  \nonumber \\
& \times   n_F^{\chi'}  (K_2) [1-n_F^{\chi} (K_3)] [1-n_F^{\chi'} (K_4)] \ , \label{eq:GammaOSEFT}
\end{align}
where the scattering amplitude in the OSEFT, at the order we computed is given in Eq.~(\ref{eq:Msq}). This scattering amplitude depends on the spin tensor of the particle, and thus on the frame vector $u^\mu$, which does not show up in the QED result, see Eq.~(\ref{ExpQEDscatt}). It is however easy to show that the OSEFT expression leads to the same value of the interaction rate as that computed with QED, at the order of accuracy we work. Using that $u^\mu =(1, {\bf 0})$ one simply has to integrate over the ${\bf K}_2$ momentum, using Eqs.~(\ref{angularint}) to reach to Eq.~(\ref{interate}), as one could naturally expect.

Finally we remind the reader that the kinetic theory expression (\ref{eq:GammaOSEFT}) can be obtained from the OSEFT as
\begin{align} 
\Gamma^\chi (E_{K}) & 
 = - \frac{1}{2} \textrm{Tr} 
\left[ P_\chi  \slashed{v} \ {\rm Im} \Sigma^R_{E,v} (X, k)\right] \ ,
\end{align}
where thanks to the cutting rules all fermions in the imaginary part of the self-energy are put on shell. In addition, the distribution functions are the equilibrium ones~(\ref{eq:eqdistro}). Instead of directly computing the retarded self-energy one could use the relation $-2 {\rm Im} \Sigma_{E,v}^R=\Sigma_{E,v}^<-\Sigma_{E,v}^> $ in the $T=0$ case, where further simplifications occur, to express $\Gamma^\chi(E_{K})$ as similar traces to those computed already in Eq.~(\ref{eq:initialtrace}).

\section{Conclusions and Outlook~\label{sec:conclusions}}

We have used the OSEFT---an effective field theory designed to describe on-shell degrees of freedom---which disentangles particles and antiparticles, for the derivation of the relativistic version of the chiral kinetic theory without and with collisions. We have also proved that the OSEFT is the quantum field theory counterpart of a Foldy-Wouthuysen diagonalization carried out for massless fermions, as opposed to the original FW approach valid for massive fermions.

The main advantage of our approach is that it has a clear semi-classical interpretation, as it is not affected by the ZB oscillations, at a given order of accuracy.  However one has to keep in mind that our effective classical particles and antiparticles have to  be viewed as combinations of the original Dirac particles and antiparticles, and they have a finite size. We have to stress that these ideas have been already implemented for massive relativistic fermions before. Our main contributions have been in generalizing them for the massless case, and use them to derive semi-classical transport equations. We have worked out an effective field theory method for that purpose, that allows us to implement this program systematically as a $1/E$ expansion. We have derived the first terms of the transport approach in this expansion, which could be pushed to higher orders if desired.

The OSEFT also allows us to understand the range of validity of the CKT, which to our knowledge, has not been discussed before. In our derivation, we have always assumed that the fermion energy is the large scale in the system. As the whole program is meant to be applied to a many-body system, this is a statement related to the {\it mean} energy of the fermions in the system. Our derivation of the CKT is valid for the fermionic modes which have energies close to that mean value, under the assumption that this is the large scale of the system. For example, close to thermal equilibrium, the mean fermion energy is of the order of the temperature $\sim T$, which is considered then to be the hard (large)  scale.  It was already realized in the past for thermal plasmas~\cite{Blaizot:2001nr,Litim:2001db}, that a pure classical transport approach is only valid for the quasiparticles with large energies, while the lower modes would not admit such a treatment.

We also note that in this semi-classical formulation, having a more accurate description of the quasiparticle hard modes including quantum effects, we end up including corrections to the classical transport equation of the order $\pa^\mu_X/E$, which then allows us to use the transport framework at  shorter distances.

In this manuscript we have presented a detailed derivation of the collision term of the chiral kinetic theory derived from the OSEFT at order $1/E$. At this order the collision term depends
on the spin tensor of the particle,  and the associated terms describe how a  massless fermion of a given chirality interacts differently with the transverse photons of different circular 
polarization in a plasma with chiral imbalance. We have checked that our collision term  for an ultradegenerate plasma, and in the presence of chiral imbalance, allows us  to reproduce the decay rate of a fermion as computed directly from QED.

Note that from the OSEFT, as initially constructed, one cannot describe fermion-antifermion annihilation processes. However, it is easy to understand that these processes are very much suppressed as compared to the particle-particle, and particle-antiparticle scatterings mediated by a soft photon exchange, only by doing a power counting analysis of their corresponding matrix elements. However, it is possible to enlarge the OSEFT to include these processes, even if they are subleading, by adding four fermion contact interactions, similarly as it is done in NRQED~\cite{Hoang:1999tp} . The resulting events would contribute to the collision term at order $1/E^4$, so that at the order we have worked in this
manuscript, they can be safely ignored.

It would be very interesting to use our results for different physical applications. In particular, for a plasma near equilibrium one could study different transport coefficients.
In the presence of chiral imbalance our framework could also be used to study the fate of the chiral plasma instabilities discussed in Ref.~\cite{Akamatsu:2013pjd}.

We have not discussed in this work the form of the transport equation for on-shell energetic photons, and how the OSEFT techniques are applied to gauge degrees of freedom. This will be the subject of future projects.

Finally, let us comment that it would also be very interesting to apply our effective field theory techniques in order to obtain semi-classical transport equations associated to massive relativistic fermions. In some recent works these equations are obtained from the Dirac picture in a $\hbar$-expansion~\cite{Weickgenannt:2019dks,Gao:2019znl,Hattori:2019ahi,Yang:2020hri,Liu:2020flb}.

\acknowledgments

J.M.T.-R. acknowledges the hospitality of the Institut de Ci\`encies de l'Espai (CSIC) and Universitat de Barcelona, where part of this work was carried out.

We have been supported by the Ministerio de Ciencia  e Innovaci\'on (Spain) under the projects FPA2016-81114-P and  FPA2016-76005-C2-1-P, as well as by the project  2017-SGR-929  (Catalonia). J.M.T.-R. acknowledges support from the Deutsche Forschungsgemeinschaft (DFG, German Research Foundation) through projects no. 411563442 (Hot Heavy Mesons) and no. 315477589 - TRR 211 (Strong-interacting matter under extreme conditions). This work was also supported by the COST Action CA15213 THOR.


\appendix

\section{Notation~\label{app:notation}}

\begin{list}{}{}
\item $E$ -- on-shell (light-like) fermion energy -- Eq.~(\ref{eq:momentum});
\item $v^\mu$ -- on-shell (light-like) fermion velocity -- Eq.~(\ref{eq:momentum});
\item $k^\mu=(k^0,{\bf k})$ -- residual 4-momentum -- Eq.~(\ref{eq:momentum});
\item $K^\mu=(K^0,{\bf K})$ -- full 4-momentum -- Eq.~(\ref{eq:momentum});
\item $E_{K}$ -- physical fermion energy -- Eq.~(\ref{eq:EK1});
\item $v^\mu_{K}$ -- on-shell (physical) velocity -- Eq.~(\ref{eq:vK1});
\item $q^\mu=(q^0,{\bf q})$ -- photon (soft) momentum -- Eq.~(\ref{eq:trace});
\item $x^\mu,y^\mu$ -- spacetime coordinates --  Eq.~(\ref{progaKS});
\item $X^\mu,s^\mu$ -- CM and relative distance coordinates --  Eq.~(\ref{eq:Wigfun});
\item $u^\mu$ -- reference frame 4-vector -- Eq.~(\ref{eq:refframe});
\item $V^{(n)}$ -- $n$-th order fermion-photon vertex in the OSEFT  -- Eq.~(\ref{eq:vertices});
\item $\Delta^k_\mu$ -- transport operator -- Eq.~(\ref{eq:LHS_G});
\item $S^< (x,y)$ -- fermion propagator in QED-- Eq.~(\ref{progaKS});
\item $S^< (X,K)$ -- fermion Wigner function in QED -- Eq.~(\ref{eq:Wigner});
\item $\Sigma (x,y)$ -- fermion self-energy in QED -- Eq.~(\ref{eq:KB1});
\item $S_{E,v} (x,y)$ -- fermion propagator in the OSEFT  -- Eq.~(\ref{eq:2pointOSEFT});
\item $S^<_{E,v} (X,k)$ -- fermion Wigner function in the OSEFT  -- Eq.~(\ref{eq:Wigfun});
\item $\Sigma_{E,v} (X,k)$ -- fermion self-energy in OSEFT -- Eq.~(\ref{eq:initialtrace});
\item $G_{E,v}^\chi (X,k)$ -- fermion 2-point function in OSEFT -- Eq.~(\ref{G-function});
\item $\Pi_{\mu \nu}^{(n)}$ -- $n$-th order photon polarization tensor -- Eq.~(\ref{eq:photonpol});
\item $D^{(n)}_{\mu 	\nu}$ -- $n$-th order photon propagator  -- Eq.~(\ref{eq:photprop});
\item ${\cal D}_L,  {\cal D}^h_T$ -- resummed longitudinal and transverse photon propagator -- Eq.~(\ref{eq:photonprop});
\item $\rho_L(q_0,{\bf q}),\rho_T^h(q_0,{\bf q})$ -- photon longitudinal and transverse spectral functions  -- Eq.~(\ref{spectral});
\item $\chi,\chi'$ -- fermion chirality -- Eq.~(\ref{eq:proj});
\item $P_\chi$ -- chiral projector -- Eq.~(\ref{eq:proj});
\item $h, h'$ -- photon helicity-- Eq.~(\ref{eq:heliproj});
\item $P_{ij}^{T,h}$ -- helicity projector -- Eq.~(\ref{eq:heliproj});
\item $P_v,P_{\tilde v}$ -- particle/antiparticle projectors -- Eq.~(\ref{eq:projectors});
\item $S^{\mu \nu}_\chi$ -- spin tensor -- Eq.~(\ref{eq:spin});
\item $\Gamma^\chi(E_K)$  -- fermion decay rate -- Eq.~(\ref{eq:decayrate});
\item $C^T,\tilde{C}^T$  -- particle-particle/particle-antiparticle total collision term -- Eq.~(\ref{eq:initialtrace}).
\end{list}



\section{The photon propagator in a medium with chiral imbalance~\label{app-photon}}

The generic form of the photon propagator in a medium where both parity $P$ and $CP$ are broken has been discussed in Ref.~\cite{Nieves:1988qz}. In a medium the photon polarization tensor has both longitudinal and transverse components, but in the presence of a chiral chemical potential, another structure antisymmetric in the Lorentz indices is also possible, respecting all the possible symmetries in the system. In turn, this implies that the photon propagator, in the Coulomb gauge for example, can be written
in terms of three components, see Eq.~(\ref{photon-propag}), a longitudinal one, and two transverse different components, which correspond to right- and left-handed circular polarizations. This is a general result, based only on the symmetries of the problem.

For a system close to thermal equilibrium, and in the presence of a chiral imbalance in the local rest frame with the plasma, it is possible to get the proper form of the photon propagator for low momenta~\cite{Carignano:2018thu}. The photon polarization tensor can actually be computed with chiral kinetic theory~\cite{Manuel:2013zaa}.

The resummed longitudinal and transverse propagators  read, with the usual prescription  $q_0 \rightarrow q_0 \pm i \eta$ for retarded and advanced quantities, respectively,
\be \label{eq:photonprop}
{\cal D}_L (q_0, \qq)= \frac{1}{\qq^2 + \Pi_L (q_0,\qq)} \ , \qquad  {\cal  D}_T^h(q_0,\qq) = \frac{1}{q_0^2 -\qq^2 - \Pi_T(q_0,\qq) - h \Pi_P (q_0,\qq)} \ ,
\ee
where 
\begin{eqnarray}
\label{pipiL}
\Pi_L (q_0,\qq) & = & m^2 _{D}  \left(1- \frac{q_0}{2 |\qq|} 
 \,{\rm ln\,}{\frac{q_0+ |\qq|}{q_0- |\qq|}}  \right)
  \ , \\
 \Pi_T (q_0,\qq) & = & m^2 _{D} \, \frac{q_0^2}{2  |\qq|^2} \left[ 1 + \frac12 \left( \frac{|\qq|}{q_0} -
\frac{q_0}{ |\qq|} \right) \,  {\rm ln\,} {\frac{q_0+
 |\qq|}{q_0- |\qq|}} 
\, \right] \ ,
 \label{pipiT}
\end{eqnarray}
are the longitudinal/transverse part of the hard thermal/dense loop photon polarization tensor \cite{LeBellac:1996kr,Manuel:2000mk},
and
 \be
 m^2_D = e^2  \left( \frac{T^2}{3} + \frac{\mu_R^2 + \mu_L^2}{2 \pi^2} \right)
\ee
is the Debye mass, 
while 
\be
\Pi_P (q_0,\qq)=  - \frac{e^2 \mu_5}{2 \pi^2}  \frac{q_0^2-|\qq|^2}{ |\qq| } \Big[ 1 - \frac{q_0}{2  |\qq|}   {\rm ln\,} {\frac{q_0+ |\qq|}{q_0-|\qq|}}
  \Big] \,
\ee  
can be viewed as the anomalous hard dense loop contribution~\cite{Laine:2005bt,Akamatsu:2013pjd,Manuel:2013zaa}.

The spectral functions associated with the gauge field modes are given by
\ba
 \rho_{L} (q_0,\qq) & =  &2 \,{\rm Im} \, {\cal D}_{L}(q_0 + i\eta, \qq) \ ,  \\
  \rho_{T}^h (q_0,\qq) & = &  2 \,{\rm Im} \, {\cal D}_{T}^h(q_0 + i\eta, \qq)  \ , \qquad h= \pm \ .
\label{spectral}
\ea

\end{document}